\documentclass[reprint,superscriptaddress, amsmath,amssymb,aps]{revtex4-1}
\usepackage{graphicx}
\pdfoutput=1
\usepackage{epstopdf}
\usepackage{color}
\usepackage{dcolumn}
\usepackage{bm}
\usepackage{hyperref}
\usepackage{upgreek} 
\usepackage{wasysym}
\usepackage{siunitx}
\usepackage{booktabs}
\usepackage[normalem]{ulem} 
\usepackage{threeparttable} 
\usepackage{tabularx}

\newcommand{\isot}[2]{$^{\textrm{#2}}$#1 }
\newcommand{\rbdoped}{$^{83}$Rb-doped }
\newcommand{\krmdoped}{$^{83\textrm{m}}$Kr-doped }
\newcommand{\rbcomma}{$^{83}$Rb, }
\newcommand{\rb}{\isot{Rb}{83}}
\newcommand{\kr}{\isot{Kr}{83}}
\newcommand{\krm}{\isot{Kr}{83m}}

\newcommand{\subStwo}{$_{\textrm{S2}}$}
\newcommand{\subreal}{$_{\textrm{vertex}}$}

\begin{document}

\title{\krm calibration of the 2013 LUX dark matter search}

\author{D.S.~Akerib} \affiliation{Case Western Reserve University, Department of Physics, 10900 Euclid Ave, Cleveland, OH 44106, USA} \affiliation{SLAC National Accelerator Laboratory, 2575 Sand Hill Road, Menlo Park, CA 94205, USA} \affiliation{Kavli Institute for Particle Astrophysics and Cosmology, Stanford University, 452 Lomita Mall, Stanford, CA 94309, USA}
\author{S.~Alsum} \affiliation{University of Wisconsin-Madison, Department of Physics, 1150 University Ave., Madison, WI 53706, USA}  
\author{H.M.~Ara\'{u}jo} \affiliation{Imperial College London, High Energy Physics, Blackett Laboratory, London SW7 2BZ, United Kingdom}  
\author{X.~Bai} \affiliation{South Dakota School of Mines and Technology, 501 East St Joseph St., Rapid City, SD 57701, USA}  
\author{A.J.~Bailey} \affiliation{Imperial College London, High Energy Physics, Blackett Laboratory, London SW7 2BZ, United Kingdom}  
\author{J.~Balajthy} \affiliation{University of Maryland, Department of Physics, College Park, MD 20742, USA}  
\author{P.~Beltrame} \affiliation{SUPA, School of Physics and Astronomy, University of Edinburgh, Edinburgh EH9 3FD, United Kingdom}  
\author{E.P.~Bernard} \affiliation{University of California Berkeley, Department of Physics, Berkeley, CA 94720, USA} \affiliation{Yale University, Department of Physics, 217 Prospect St., New Haven, CT 06511, USA} 
\author{A.~Bernstein} \affiliation{Lawrence Livermore National Laboratory, 7000 East Ave., Livermore, CA 94551, USA}  
\author{T.P.~Biesiadzinski} \affiliation{Case Western Reserve University, Department of Physics, 10900 Euclid Ave, Cleveland, OH 44106, USA} \affiliation{SLAC National Accelerator Laboratory, 2575 Sand Hill Road, Menlo Park, CA 94205, USA} \affiliation{Kavli Institute for Particle Astrophysics and Cosmology, Stanford University, 452 Lomita Mall, Stanford, CA 94309, USA}
\author{E.M.~Boulton} \affiliation{University of California Berkeley, Department of Physics, Berkeley, CA 94720, USA} \affiliation{Yale University, Department of Physics, 217 Prospect St., New Haven, CT 06511, USA} 
\author{P.~Br\'as} \affiliation{LIP-Coimbra, Department of Physics, University of Coimbra, Rua Larga, 3004-516 Coimbra, Portugal}  
\author{D.~Byram} \affiliation{University of South Dakota, Department of Physics, 414E Clark St., Vermillion, SD 57069, USA} \affiliation{South Dakota Science and Technology Authority, Sanford Underground Research Facility, Lead, SD 57754, USA} 
\author{S.B.~Cahn} \affiliation{Yale University, Department of Physics, 217 Prospect St., New Haven, CT 06511, USA}  
\author{M.C.~Carmona-Benitez} \affiliation{Pennsylvania State University, Department of Physics, 104 Davey Lab, University Park, PA  16802-6300, USA} \affiliation{University of California Santa Barbara, Department of Physics, Santa Barbara, CA 93106, USA} 
\author{C.~Chan} \affiliation{Brown University, Department of Physics, 182 Hope St., Providence, RI 02912, USA}  
\author{A.~Currie} \affiliation{Imperial College London, High Energy Physics, Blackett Laboratory, London SW7 2BZ, United Kingdom}  
\author{J.E.~Cutter} \affiliation{University of California Davis, Department of Physics, One Shields Ave., Davis, CA 95616, USA}  
\author{T.J.R.~Davison} \affiliation{SUPA, School of Physics and Astronomy, University of Edinburgh, Edinburgh EH9 3FD, United Kingdom}  
\author{A.~Dobi} \affiliation{Lawrence Berkeley National Laboratory, 1 Cyclotron Rd., Berkeley, CA 94720, USA}  
\author{E.~Druszkiewicz} \affiliation{University of Rochester, Department of Physics and Astronomy, Rochester, NY 14627, USA}  
\author{B.N.~Edwards} \affiliation{Yale University, Department of Physics, 217 Prospect St., New Haven, CT 06511, USA}  
\author{S.R.~Fallon} \affiliation{University at Albany, State University of New York, Department of Physics, 1400 Washington Ave., Albany, NY 12222, USA}  
\author{A.~Fan} \affiliation{SLAC National Accelerator Laboratory, 2575 Sand Hill Road, Menlo Park, CA 94205, USA} \affiliation{Kavli Institute for Particle Astrophysics and Cosmology, Stanford University, 452 Lomita Mall, Stanford, CA 94309, USA} 
\author{S.~Fiorucci} \affiliation{Lawrence Berkeley National Laboratory, 1 Cyclotron Rd., Berkeley, CA 94720, USA} \affiliation{Brown University, Department of Physics, 182 Hope St., Providence, RI 02912, USA} 
\author{R.J.~Gaitskell} \affiliation{Brown University, Department of Physics, 182 Hope St., Providence, RI 02912, USA}  
\author{J.~Genovesi} \affiliation{University at Albany, State University of New York, Department of Physics, 1400 Washington Ave., Albany, NY 12222, USA}  
\author{C.~Ghag} \affiliation{Department of Physics and Astronomy, University College London, Gower Street, London WC1E 6BT, United Kingdom}  
\author{M.G.D.~Gilchriese} \affiliation{Lawrence Berkeley National Laboratory, 1 Cyclotron Rd., Berkeley, CA 94720, USA}  
\author{C.R.~Hall} \affiliation{University of Maryland, Department of Physics, College Park, MD 20742, USA}  
\author{M.~Hanhardt} \affiliation{South Dakota School of Mines and Technology, 501 East St Joseph St., Rapid City, SD 57701, USA} \affiliation{South Dakota Science and Technology Authority, Sanford Underground Research Facility, Lead, SD 57754, USA} 
\author{S.J.~Haselschwardt} \affiliation{University of California Santa Barbara, Department of Physics, Santa Barbara, CA 93106, USA}  
\author{S.A.~Hertel} 
\email{scott.hertel@umass.edu}\affiliation{University of Massachusetts, Amherst Center for Fundamental Interactions and Department of Physics, Amherst, MA 01003-9337 USA} \affiliation{Lawrence Berkeley National Laboratory, 1 Cyclotron Rd., Berkeley, CA 94720, USA} \affiliation{Yale University, Department of Physics, 217 Prospect St., New Haven, CT 06511, USA}
\author{D.P.~Hogan} \affiliation{University of California Berkeley, Department of Physics, Berkeley, CA 94720, USA}  
\author{M.~Horn} \affiliation{South Dakota Science and Technology Authority, Sanford Underground Research Facility, Lead, SD 57754, USA} \affiliation{University of California Berkeley, Department of Physics, Berkeley, CA 94720, USA} \affiliation{Yale University, Department of Physics, 217 Prospect St., New Haven, CT 06511, USA}
\author{D.Q.~Huang} \affiliation{Brown University, Department of Physics, 182 Hope St., Providence, RI 02912, USA}  
\author{C.M.~Ignarra} \affiliation{SLAC National Accelerator Laboratory, 2575 Sand Hill Road, Menlo Park, CA 94205, USA} \affiliation{Kavli Institute for Particle Astrophysics and Cosmology, Stanford University, 452 Lomita Mall, Stanford, CA 94309, USA} 
\author{R.G.~Jacobsen} \affiliation{University of California Berkeley, Department of Physics, Berkeley, CA 94720, USA}  
\author{W.~Ji} \affiliation{Case Western Reserve University, Department of Physics, 10900 Euclid Ave, Cleveland, OH 44106, USA} \affiliation{SLAC National Accelerator Laboratory, 2575 Sand Hill Road, Menlo Park, CA 94205, USA} \affiliation{Kavli Institute for Particle Astrophysics and Cosmology, Stanford University, 452 Lomita Mall, Stanford, CA 94309, USA}
\author{K.~Kamdin} \affiliation{University of California Berkeley, Department of Physics, Berkeley, CA 94720, USA}  
\author{K.~Kazkaz} \affiliation{Lawrence Livermore National Laboratory, 7000 East Ave., Livermore, CA 94551, USA}  
\author{D.~Khaitan} \affiliation{University of Rochester, Department of Physics and Astronomy, Rochester, NY 14627, USA}  
\author{R.~Knoche} \affiliation{University of Maryland, Department of Physics, College Park, MD 20742, USA}  
\author{N.A.~Larsen} \affiliation{Yale University, Department of Physics, 217 Prospect St., New Haven, CT 06511, USA}  
\author{B.G.~Lenardo} \affiliation{University of California Davis, Department of Physics, One Shields Ave., Davis, CA 95616, USA} \affiliation{Lawrence Livermore National Laboratory, 7000 East Ave., Livermore, CA 94551, USA} 
\author{K.T.~Lesko} \affiliation{Lawrence Berkeley National Laboratory, 1 Cyclotron Rd., Berkeley, CA 94720, USA}  
\author{A.~Lindote} \affiliation{LIP-Coimbra, Department of Physics, University of Coimbra, Rua Larga, 3004-516 Coimbra, Portugal}  
\author{M.I.~Lopes} \affiliation{LIP-Coimbra, Department of Physics, University of Coimbra, Rua Larga, 3004-516 Coimbra, Portugal}  
\author{A.~Manalaysay} \affiliation{University of California Davis, Department of Physics, One Shields Ave., Davis, CA 95616, USA}  
\author{R.L.~Mannino} \affiliation{Texas A \& M University, Department of Physics, College Station, TX 77843, USA} \affiliation{University of Wisconsin-Madison, Department of Physics, 1150 University Ave., Madison, WI 53706, USA} 
\author{M.F.~Marzioni} \affiliation{SUPA, School of Physics and Astronomy, University of Edinburgh, Edinburgh EH9 3FD, United Kingdom}  
\author{D.N.~McKinsey} \affiliation{University of California Berkeley, Department of Physics, Berkeley, CA 94720, USA} \affiliation{Lawrence Berkeley National Laboratory, 1 Cyclotron Rd., Berkeley, CA 94720, USA} \affiliation{Yale University, Department of Physics, 217 Prospect St., New Haven, CT 06511, USA}
\author{D.-M.~Mei} \affiliation{University of South Dakota, Department of Physics, 414E Clark St., Vermillion, SD 57069, USA}  
\author{J.~Mock} \affiliation{University at Albany, State University of New York, Department of Physics, 1400 Washington Ave., Albany, NY 12222, USA}  
\author{M.~Moongweluwan} \affiliation{University of Rochester, Department of Physics and Astronomy, Rochester, NY 14627, USA}  
\author{J.A.~Morad} \affiliation{University of California Davis, Department of Physics, One Shields Ave., Davis, CA 95616, USA}  
\author{A.St.J.~Murphy} \affiliation{SUPA, School of Physics and Astronomy, University of Edinburgh, Edinburgh EH9 3FD, United Kingdom}  
\author{C.~Nehrkorn} \affiliation{University of California Santa Barbara, Department of Physics, Santa Barbara, CA 93106, USA}  
\author{H.N.~Nelson} \affiliation{University of California Santa Barbara, Department of Physics, Santa Barbara, CA 93106, USA}  
\author{F.~Neves} \affiliation{LIP-Coimbra, Department of Physics, University of Coimbra, Rua Larga, 3004-516 Coimbra, Portugal}  
\author{K.~O'Sullivan} \affiliation{University of California Berkeley, Department of Physics, Berkeley, CA 94720, USA} \affiliation{Lawrence Berkeley National Laboratory, 1 Cyclotron Rd., Berkeley, CA 94720, USA} \affiliation{Yale University, Department of Physics, 217 Prospect St., New Haven, CT 06511, USA}
\author{K.C.~Oliver-Mallory} \affiliation{University of California Berkeley, Department of Physics, Berkeley, CA 94720, USA}  
\author{K.J.~Palladino} \affiliation{University of Wisconsin-Madison, Department of Physics, 1150 University Ave., Madison, WI 53706, USA} \affiliation{SLAC National Accelerator Laboratory, 2575 Sand Hill Road, Menlo Park, CA 94205, USA} \affiliation{Kavli Institute for Particle Astrophysics and Cosmology, Stanford University, 452 Lomita Mall, Stanford, CA 94309, USA}
\author{E.K.~Pease} \affiliation{University of California Berkeley, Department of Physics, Berkeley, CA 94720, USA} \affiliation{Yale University, Department of Physics, 217 Prospect St., New Haven, CT 06511, USA} 
\author{C.~Rhyne} \affiliation{Brown University, Department of Physics, 182 Hope St., Providence, RI 02912, USA}  
\author{S.~Shaw} \affiliation{University of California Santa Barbara, Department of Physics, Santa Barbara, CA 93106, USA} \affiliation{Department of Physics and Astronomy, University College London, Gower Street, London WC1E 6BT, United Kingdom} 
\author{T.A.~Shutt} \affiliation{Case Western Reserve University, Department of Physics, 10900 Euclid Ave, Cleveland, OH 44106, USA}  \affiliation{Kavli Institute for Particle Astrophysics and Cosmology, Stanford University, 452 Lomita Mall, Stanford, CA 94309, USA}
\author{C.~Silva} \affiliation{LIP-Coimbra, Department of Physics, University of Coimbra, Rua Larga, 3004-516 Coimbra, Portugal}  
\author{M.~Solmaz} \affiliation{University of California Santa Barbara, Department of Physics, Santa Barbara, CA 93106, USA}  
\author{V.N.~Solovov} \affiliation{LIP-Coimbra, Department of Physics, University of Coimbra, Rua Larga, 3004-516 Coimbra, Portugal}  
\author{P.~Sorensen} \affiliation{Lawrence Berkeley National Laboratory, 1 Cyclotron Rd., Berkeley, CA 94720, USA}  
\author{T.J.~Sumner} \affiliation{Imperial College London, High Energy Physics, Blackett Laboratory, London SW7 2BZ, United Kingdom}  
\author{M.~Szydagis} \affiliation{University at Albany, State University of New York, Department of Physics, 1400 Washington Ave., Albany, NY 12222, USA}  
\author{D.J.~Taylor} \affiliation{South Dakota Science and Technology Authority, Sanford Underground Research Facility, Lead, SD 57754, USA}  
\author{W.C.~Taylor} \affiliation{Brown University, Department of Physics, 182 Hope St., Providence, RI 02912, USA}  
\author{B.P.~Tennyson} \affiliation{Yale University, Department of Physics, 217 Prospect St., New Haven, CT 06511, USA}  
\author{P.A.~Terman} \affiliation{Texas A \& M University, Department of Physics, College Station, TX 77843, USA}  
\author{D.R.~Tiedt} \affiliation{South Dakota School of Mines and Technology, 501 East St Joseph St., Rapid City, SD 57701, USA}  
\author{W.H.~To} \affiliation{California State University Stanislaus, Department of Physics, 1 University Circle, Turlock, CA 95382, USA} \affiliation{SLAC National Accelerator Laboratory, 2575 Sand Hill Road, Menlo Park, CA 94205, USA} \affiliation{Kavli Institute for Particle Astrophysics and Cosmology, Stanford University, 452 Lomita Mall, Stanford, CA 94309, USA}
\author{M.~Tripathi} \affiliation{University of California Davis, Department of Physics, One Shields Ave., Davis, CA 95616, USA}  
\author{L.~Tvrznikova} \affiliation{University of California Berkeley, Department of Physics, Berkeley, CA 94720, USA} \affiliation{Yale University, Department of Physics, 217 Prospect St., New Haven, CT 06511, USA} 
\author{S.~Uvarov} \affiliation{University of California Davis, Department of Physics, One Shields Ave., Davis, CA 95616, USA}  
\author{V.~Velan} \affiliation{University of California Berkeley, Department of Physics, Berkeley, CA 94720, USA}  
\author{J.R.~Verbus} \affiliation{Brown University, Department of Physics, 182 Hope St., Providence, RI 02912, USA}  
\author{R.C.~Webb} \affiliation{Texas A \& M University, Department of Physics, College Station, TX 77843, USA}  
\author{J.T.~White} \affiliation{Texas A \& M University, Department of Physics, College Station, TX 77843, USA}  
\author{T.J.~Whitis} \affiliation{Case Western Reserve University, Department of Physics, 10900 Euclid Ave, Cleveland, OH 44106, USA} \affiliation{SLAC National Accelerator Laboratory, 2575 Sand Hill Road, Menlo Park, CA 94205, USA} \affiliation{Kavli Institute for Particle Astrophysics and Cosmology, Stanford University, 452 Lomita Mall, Stanford, CA 94309, USA}
\author{M.S.~Witherell} \affiliation{Lawrence Berkeley National Laboratory, 1 Cyclotron Rd., Berkeley, CA 94720, USA}  
\author{F.L.H.~Wolfs} \affiliation{University of Rochester, Department of Physics and Astronomy, Rochester, NY 14627, USA}  
\author{J.~Xu} \affiliation{Lawrence Livermore National Laboratory, 7000 East Ave., Livermore, CA 94551, USA}  
\author{K.~Yazdani} \affiliation{Imperial College London, High Energy Physics, Blackett Laboratory, London SW7 2BZ, United Kingdom}  
\author{S.K.~Young} \affiliation{University at Albany, State University of New York, Department of Physics, 1400 Washington Ave., Albany, NY 12222, USA}  
\author{C.~Zhang} \affiliation{University of South Dakota, Department of Physics, 414E Clark St., Vermillion, SD 57069, USA}

\collaboration{LUX Collaboration}

\date{\today}
\vspace{10 mm}
\begin{abstract}
LUX was the first dark matter experiment to use a \krm calibration source. In this paper we describe the source preparation and injection.  We also present several \krm calibration applications in the context of the 2013 LUX exposure, including the measurement of temporal and spatial variation in scintillation and charge signal amplitudes, and several methods to understand the electric field within the time projection chamber. 
\end{abstract}

\maketitle

\section{\krm as a calibration source}

The LUX experiment searches for galactic dark matter particles scattering on target nuclei in a dual-phase xenon time projection chamber (TPC). Energy depositions in the liquid Xe (LXe) produce observable signals via prompt scintillation (S1) and ionization charge, where liberated electrons drift upwards in an applied electric field and generate a delayed electroluminescence signal (S2) in the gaseous Xe (GXe). Light from both S1 and S2 is detected by photomultiplier tubes (PMTs) situated in two 61-PMT arrays above and below the 250\,kg active xenon mass (see Ref.\,\cite{luxdetector} for more details on detector design). The energy of an event may be inferred from the amplitude of its S1 and S2 signals. Additionally, and of vital importance in rejecting background events, the 3D position of an interaction may also be reconstructed. From the S2 signal, the distribution of photons in the top PMT array localizes the event in the $xy$-plane. The $z$ position is calculated from the ionization electron drift time, i.e., the time interval separating the S1 and S2 signals.

\begin{figure}[t!]
\includegraphics[width=80mm]{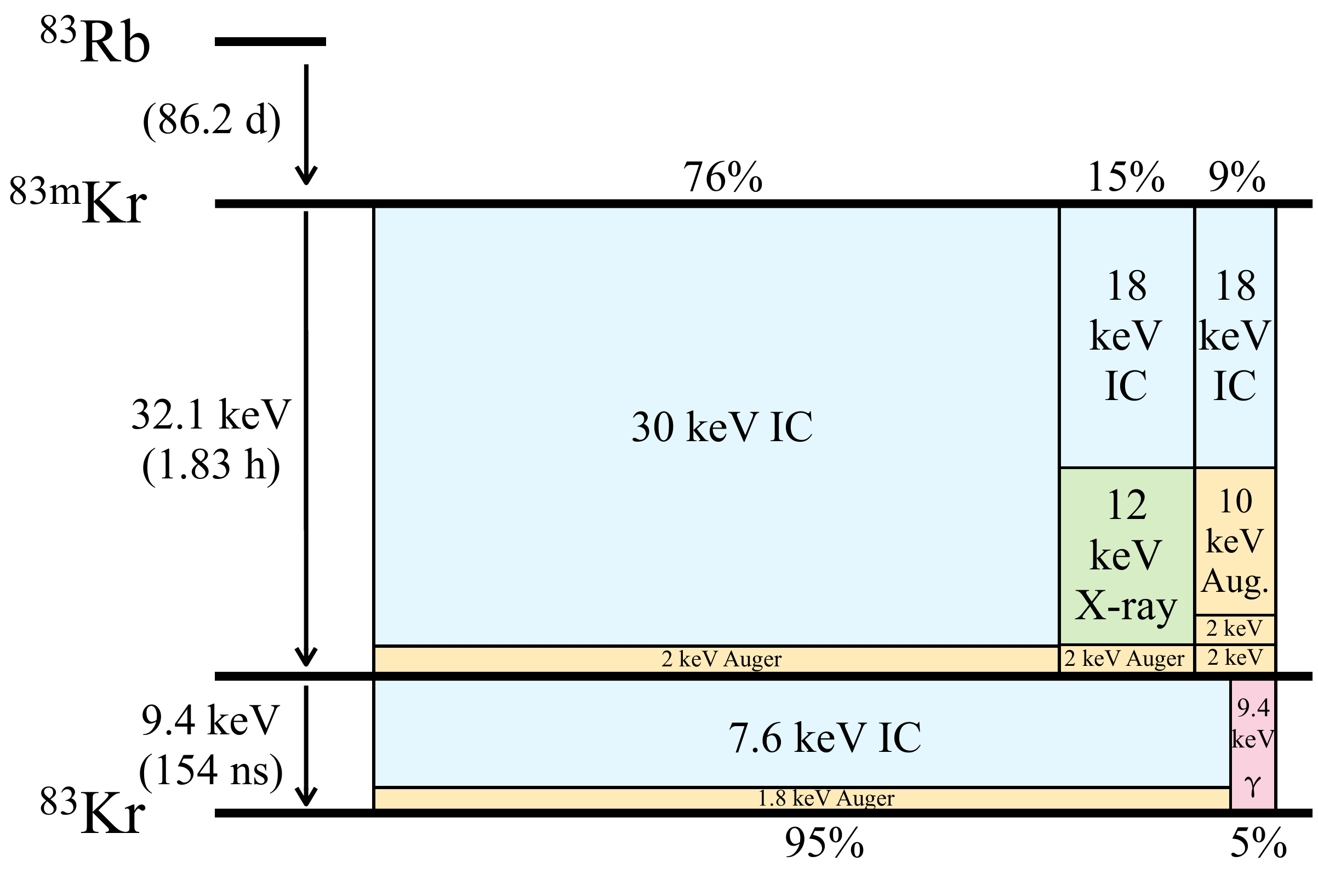}
\caption{Decay scheme of \krm.  The width of each column is proportional to the branching fraction of that decay mode, the vertical divisions are proportional to energy partitioning among internal conversion electrons (blue), Auger electrons (yellow), x-rays (green), and gamma-rays (red).  Numerical values from Reference~\cite{delphi}.}
\label{fig:decay_diagram}
\end{figure}

LUX has made extensive use of \krm for calibration purposes. The decay of \krm is illustrated in Figure~\ref{fig:decay_diagram}.  The parent isotope \rb is a practical source of \krm, due in part to its long half-life of 86.2~d. Once produced, the noble gas \krm may diffuse from the generator material into the detector volume, decaying to \kr with a half-life of 1.83~h, and releasing a total energy of 41.5~keV.  The decay occurs in two transitions of 32.1 and 9.4~keV respectively, with an intervening half-life of 154~ns. These two transitions can each proceed according to multiple decay channels as indicated in Figure~\ref{fig:decay_diagram}, but in summary \krm exhibits a high probability of internal conversion (IC) followed by Auger emission, resulting in the high concentration of decay energy into electron modes.  Two lower-probability modes of photon (gamma or x-ray) emission can occur, with a maximum photon energy of 12~keV.

The first uses of \krm as a calibration source were in ALEPH~\cite{aleph} and DELPHI~\cite{delphi}, with subsequent deployments at STAR~\cite{star} and ALICE~\cite{alice}. The IC and Auger electrons have served individually as electron energy calibration lines in experiments measuring the tritium spectrum at its endpoint (Mainz~\cite{kr83m_decay_measurement}, Triotsk~\cite{Triotsk}, KATRIN~\cite{katrin}, Project~8~\cite{project8}). \krm is a natural choice for calibrating liquid noble-element dark matter direct detection experiments due to its inert nature and keV-scale decay energy, similar to the energy scales sensed by these experiments. Initial demonstrations of \krm calibration in liquid xenon, liquid argon, and liquid neon were  performed at Yale University~\cite{kastens, rbbinding, lippincott}. The LXe response of \krm has since been studied in detail, including \cite{manalaysay2010} and \cite{aprile2012}. It has been used as a calibration source for liquid argon detectors by the SCENE collaboration~\cite{Alexander, Cao}, and to characterize a cryogenic distillation system~\cite{Rosendahl}.

In a liquid noble environment, the low-energy electrons and photons released by the decay deposit their energy within $\mathcal{O}(10~\mu\textrm{m})$ of the decay vertex.  These separations are much smaller than the spatial resolving power of the LUX detector ($\mathcal{O}(1~\textrm{mm})$~\cite{mercury}) or the typical electron diffusion distances during drift (also $\mathcal{O}(1~\textrm{mm})$~\cite{electrondiffusion}).  

We describe here the first use of \krm to directly calibrate a dark matter experiment. This paper describes the use of \krm during the first (2013) exposure of the LUX experiment~\cite{lux2013,lux2015}. The Darkside-50~\cite{darkside2016} and XENON1T~\cite{XENON1T} collaborations have reported similar calibrations.

\section{\krm hardware and mixing}

Brookhaven National Laboratory produced the \rb for LUX, via proton irradiation of a $^{\textrm{nat}}$RbCl target. Additional Rb radioisotopes can be produced, but with lower efficiency and shorter half-lives ($^{86}$Rb 18.7d, $^{84}$Rb 32.9d). The resulting \rb is stored in aqueous solution for distribution. After dilution to reduce the specific activity, a measured volume of the \rb solution is deposited on several grams of activated coconut carbon mediator (Calgon OVC 4x8). The carbon is baked at $\sim$100$^\circ$~C for several hours under vacuum, to remove water and any other volatiles.  This charcoal mediator was selected for its low radon emanation rate, previously measured to be 9.4~mBq/kg~\cite{radon}. Previous studies have found excellent binding of \rb to charcoal mediators~\cite{rbbinding}.

\begin{figure}[t!]
\includegraphics[width=80mm]{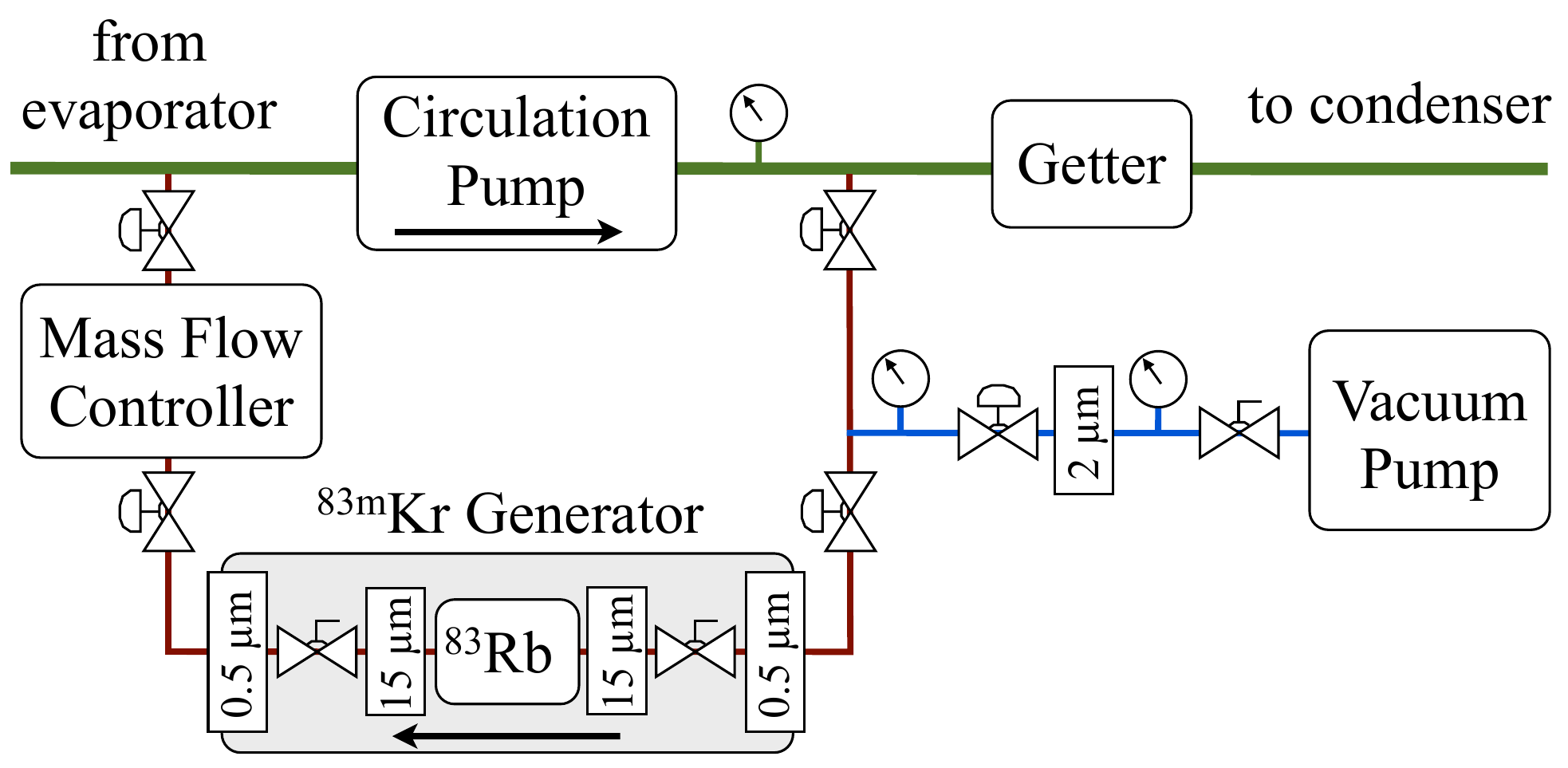}
\caption{Simplified plumbing and instrumentation diagram showing the \krm generator (grey background) and its setting for controlled injection.  The injection path (red) starts at a high-pressure point on the main Xe circulation path (green) and ends at a low-pressure point near the main circulation pump inlet.  A vacuum pump and its associated pump-out line (blue) is used to evacuate the \krm generator in some injection sequences.  Valves with semicircle handles are automated, all others are manual.  Several particulate filters are noted, labeled by their pore diameter.  Pressure gauges which play a role in the automated injection script are indicated by circles.}
\label{fig:PandID}
\end{figure}

The \rbdoped mediator is installed in the injection plumbing, as illustrated in Figure~\ref{fig:PandID}.  To prevent the spreading of possible charcoal particulates, the \rbdoped mediator is contained between two sets of particulate filtering, with pore size 15~$\mu$m and 0.5~$\mu$m. The \krm generator plumbing straddles a pressure differential in the main LUX gaseous Xe (GXe) circulation path.  During injection this pressure differential motivates flow of GXe over the mediator and into circulation.  The pressure differential is produced by the main GXe circulation pump, and the rate of GXe flow through the \krm generator is controlled using a mass flow controller downstream from the mediator, with a typical control value of 0.50~slpm (much smaller than the flow of the main circulation path).  The \krmdoped GXe passes through a getter (SAES MonoTorr~\cite{getter}) containing a 3~nm filter, further mitigating the risk of particulate contamination, or non-noble radioisotope contamination (including by atomic \rb) of the detector volume.

To release \krm calibration doses of the desired activity and duration, the \krm injection system was operated in two modes, depending on the \rb activity on the date of injection.  For low-activity \rbcomma the valves along the injection flow path (red in Figure~\ref{fig:PandID}) were simply opened for a duration proportional to the desired \krm dose, typically several minutes.  For high-activity \rbcomma the \krm generator volume was initially pumped to vacuum to eliminate the relic \krm activity prior to injection.  In this mode, the injected activity resulted only from \rb decays that occur during the injection time window (again an easily-controlled timescale, on the order of minutes).

\begin{figure}[t!]
\includegraphics[width=80mm]{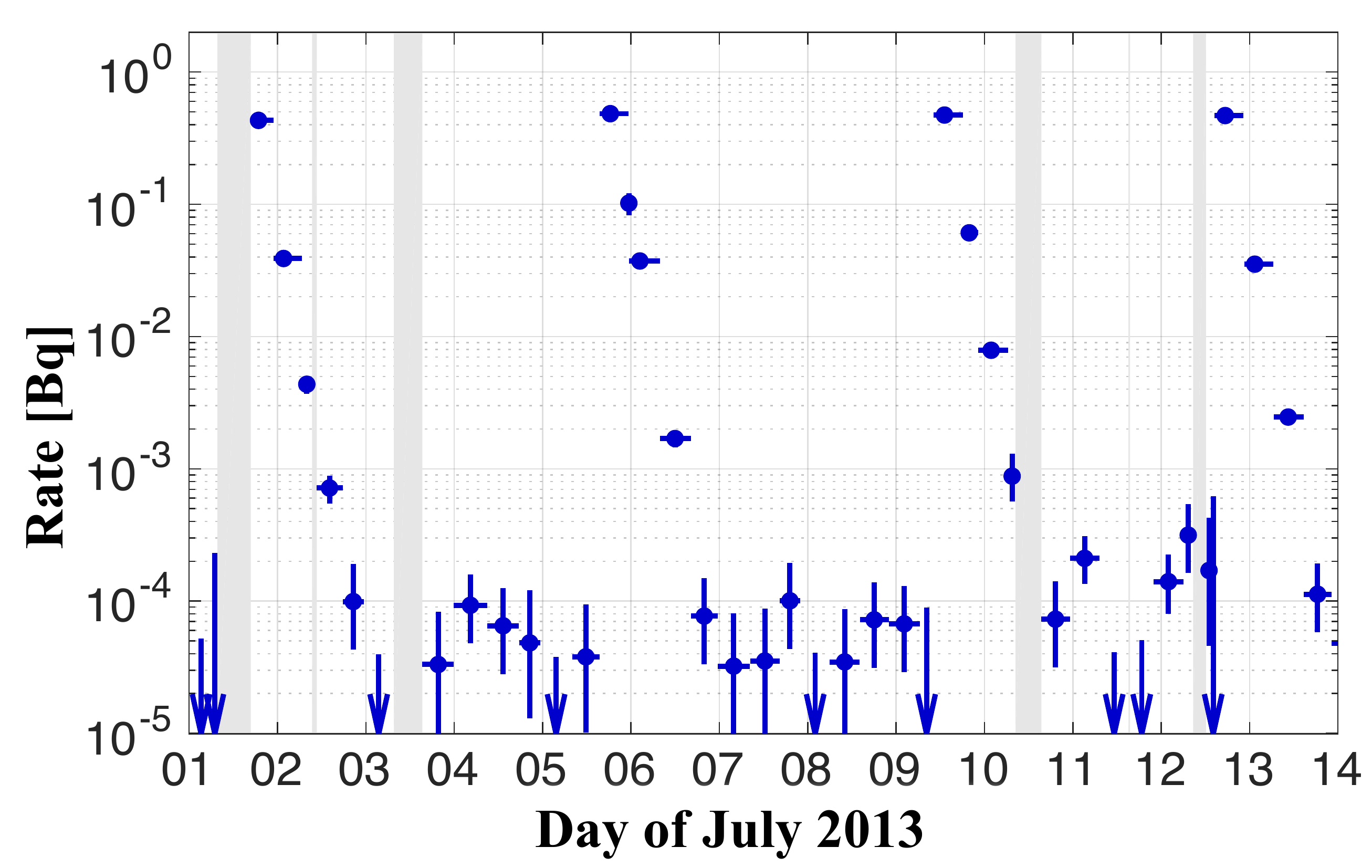}
\caption{The \krm rate within a fiducial volume selection over a period of two weeks, during which four injections were performed. The dosing system is able to inject a small and repeatable activity. For small injections, it takes $\sim$1~day for the \krm activity to fall below the baseline electron recoil background rate for that energy.}
\label{fig:repetition}
\end{figure}

Calibrations using \krm were performed on a regular (typically weekly) schedule throughout the data-taking campaign. An example of precise regular dosing is shown in Figure~\ref{fig:repetition}. A typical activity of $\sim$10~Bq was optimal for the measurement of electron lifetime in of LXe (see Section~\ref{areacorrections}), but depending on the specific calibration goal, both higher- ($\sim$100~Bq) and lower-rate injections were also performed. Hardware interlocks on pressure and flow readings would abort the injection in the event of unusual readings.

\begin{figure}[t!]
\includegraphics[width=80mm]{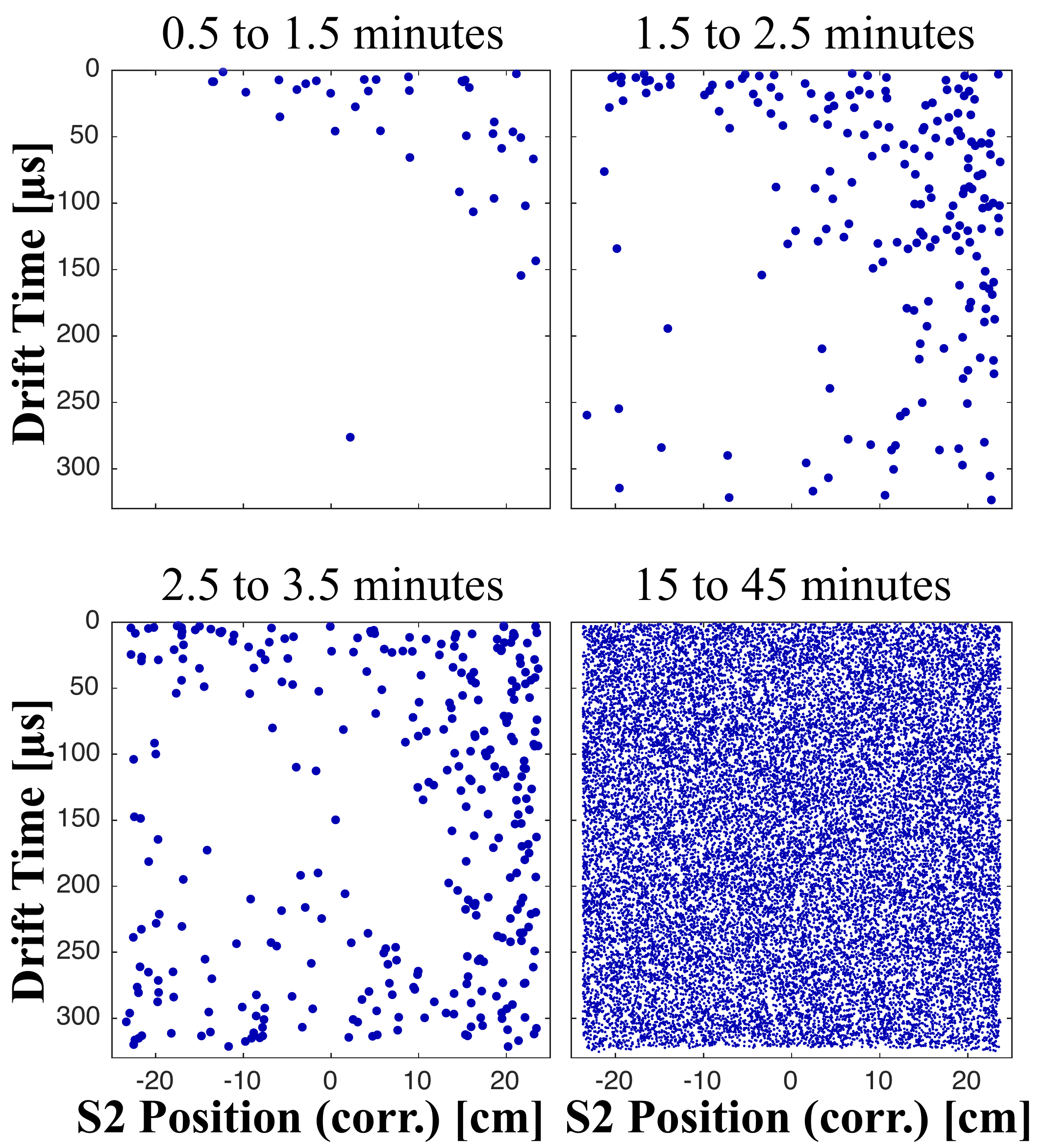}
\caption{Reconstructed \krm vertex positions are illustrated within a thin slice of the LUX TPC for four distinct time windows after a \krm injection.  The $x$-axis is the S2 $xy$ position and the $y$-axis the electron drift time as measured by the time delay between S1 and S2 (the liquid surface is at zero drift time).  The $x$-axis has here been rotated 45 degrees with respect to the typical LUX convention to better align with the dodecagonal shape of the TPC and the observed LXe flow axis.  We use here the `corrected' $xy$ coordinates as described in Section\ref{poscorr}. A large-scale flow (clockwise in these coordinates) is observed, along with turbulent mixing.}
\label{fig:mixing1}
\end{figure}

\begin{figure}[t!]
\includegraphics[width=80mm]{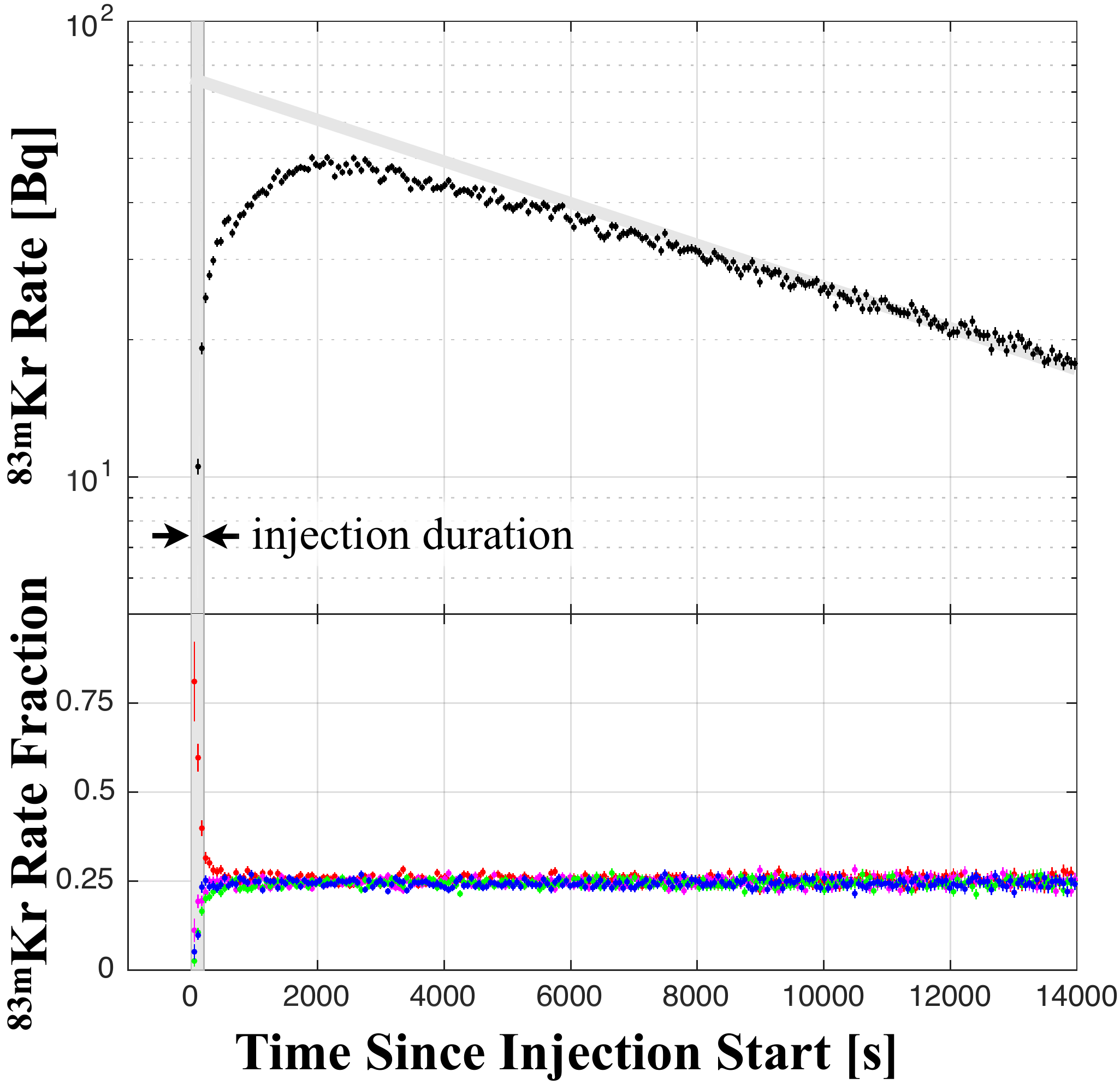}
\caption{TOP: The rate of \krm decays within the fiducial volume is shown as a function of time since the injection start for a typical high-rate injection.  The injection time (time during which GXe is flowing over the \rb parent) is indicated by a gray band, in this case lasting several minutes.  An exponential decay expressing the \krm decay half-life of 1.83h is highlighted in gray.  BOTTOM: \krm decays are here selected by drift time into four approximately equal volumes (red indicating the top fourth), and the relative fraction of \krm activity in each fourth (R$_{\textrm{1/4}}$/R$_{\textrm{tot}}$) is plotted.  The strong LXe mixing produces a homogeneous \krm activity only minutes after injection, despite the gradual arrival of \krm over one hour.}
\label{fig:mixing2}
\end{figure}

The flow and mixing of LXe within the LUX time-projection chamber (TPC) can be observed using the \krm injections, as illustrated in Figures~\ref{fig:mixing1} and \ref{fig:mixing2}.  Starting 60 seconds after GXe flow was initiated over the \rbdoped mediator, the first \krm decays are seen near the liquid surface.  A LXe flow (likely convective in origin) circulates this \krmdoped liquid with a velocity of a few cm/s, completing a circuit from top to bottom and back over $\sim$2 minutes.  As seen in the last panel of Figure~\ref{fig:mixing1} and the lower panel of Figure~\ref{fig:mixing2}, \krm activity is uniformly distributed after several minutes. We assume the activity is spatially homogeneous once the \krm distribution is observed to be constant.  The LXe flow pattern observed with \krm was consistent with similar flow observations using $^{222}$Rn-$^{218}$Po delayed $\alpha$-particle coincidences~\cite{mallingthesis}.

As shown in the upper panel of Figure~\ref{fig:mixing2}, a low rate of \krm continues to enter the detector one hour after the injection sequence.  This is attributed to \krm activity slowly diffusing out the long and narrow GXe volume between the circulation path and the last outlet valve of the \krm injection line.

S1 and S2 pulse amplitudes for \krm decays lie outside the typical dark matter search window.  Further, the presence of \krm activity was seen to not increase the rate of low-energy triggers passing selection criteria applied as in~\cite{lux2015}.  While it appears then that \krm activity does not necessarily disqualify data from a low-energy low-background search, the short half-life allows the conservative exclusion of this data with no significant decrease in search exposure.

In the following sections, we describe the use of \krm for several calibrations central to the 2013 dark matter search of LUX, a search summarized in references \cite{lux2013} and \cite{lux2015}.

\section{Studies of the electric field}\label{fieldmodel}

The 3D position reconstruction of ionization vertices requires an understanding of the path electrons take from their production site to their point of detection. In LUX, the latter occurs in close proximity to the liquid surface, where the observed S2 signal is generated via electroluminescence in GXe at high field.  The distribution of S2 light sensed by the top PMT array is converted into an S2 position ($x$\subStwo, $y$\subStwo) using algorithms described in~\cite{mercury}. While the electric field is largely perpendicular to the liquid surface at all positions, the electric field lines in the first LUX science run (WS2013) include a small but non-zero radial component, inducing a radially-inward electron drift.  This radial field component is due to the non-zero electrostatic transparency of the field cage. \krm calibrations fill the TPC to its edges with a uniform specific activity (activity per unit LXe volume), allowing for a robust consistency check of the observed drift field with that expected from geometrical effects alone.

\begin{figure}[t!]
\includegraphics[width=80mm]{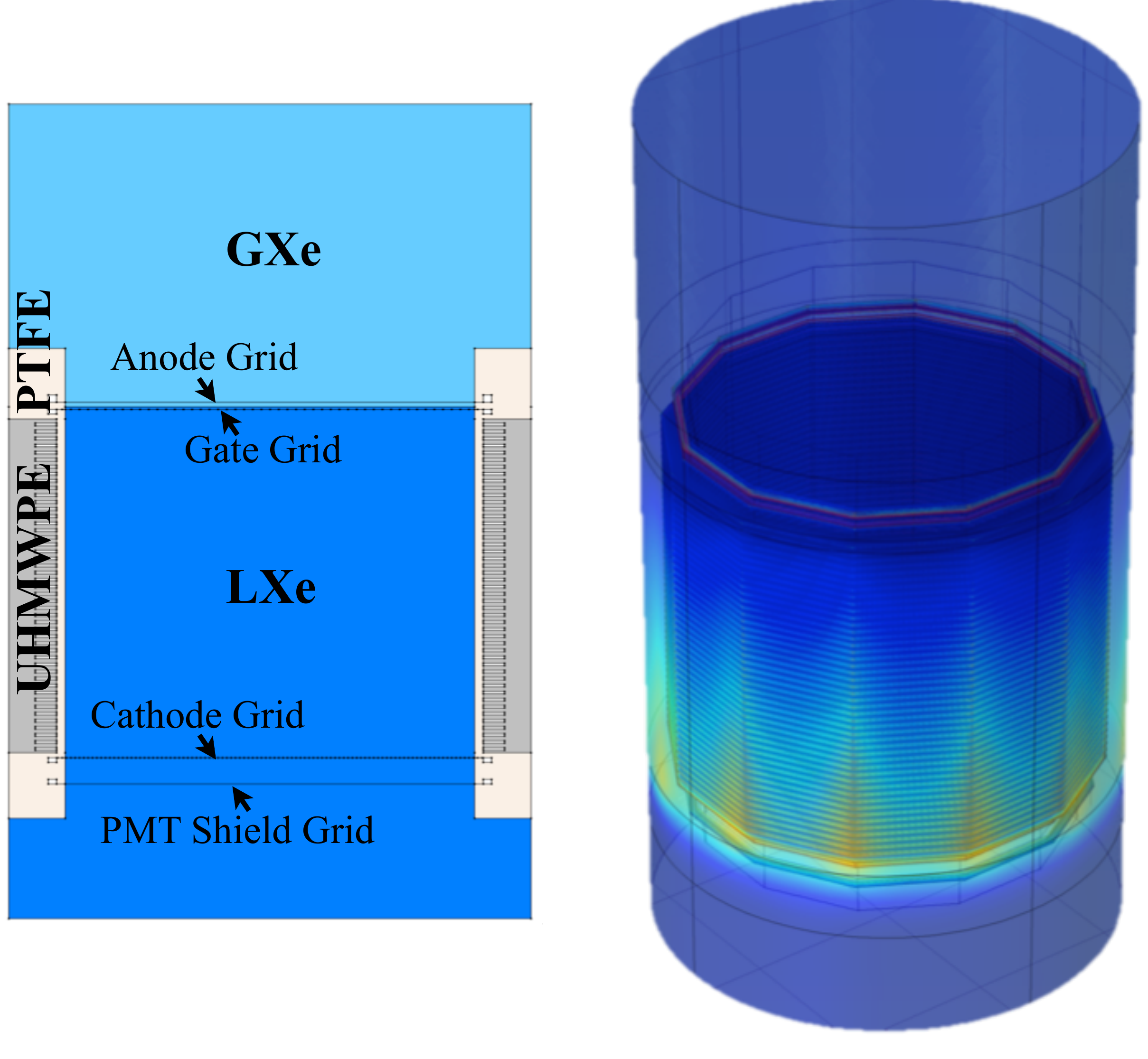}
\caption{LEFT: Illustration of key features of the 3D LUX model, labeling materials and grids. The model is bounded by a the inner radius of the cryostat inner vessel at 31 cm.  The central volume of LXe is bounded by 12 PTFE panels each of width 12.7~cm, forming a dodecagon of 23.7~cm apothem (radius of inscribed circle). As described in the text, the anode and bottom PMT shield grids are modeled as solid planes (making inclusion of detailed model geometry above the anode and below the bottom PMT shield unnecessary). RIGHT: A 3D map of electric field is obtained after the model in COMSOL is built, meshed and solved. Note the dodecahedral symmetry of the model in the relevant region.} \label{fig:geometry} \end{figure}

A 3D model of the LUX geometry is constructed in COMSOL Multiphysics\textsuperscript{\textregistered}~\cite{comsol}. A 2D cross-section of this model is shown in Figure~\ref{fig:geometry}. Due to the detector's geometrical complexity (relevant dimensions span 4 orders of magnitude), several model simplifications are adopted, each of which has been checked to ensure the simplification is of negligible effect to the resulting drift field.  Details of boundaries within the ultra-high-molecular-weight polyethylene (UHMWPE) and polytetrafluoroethylene (PTFE) volumes are omitted, including the weir, cathode cable and the heat exchanger. The anode grid and the bottom photomultiplier tube (PMT) shield grid are both modeled not as wires but as planes (the anode grid wires are of sub-mm spacing, and the bottom shield grid is backed by PMT faces of similar voltage). The cathode and gate grids are accurately modeled as parallel wires of appropriate spacing, thereby accounting for the electrostatic transparency of the real detector grids. These cathode and gate grids are simplified only in that the wire diameter is reduced to zero (from 206 and 101.6~$\mu$m respectively).  Test models were studied to ensure this wire diameter change had negligible effect on the resulting solution, as expected from COMSOL's use of the weak formulation~\cite{weakform} in solving the relevant partial differential equations.

The TPC diameter as measured between parallel opposite faces is 47.3~cm. The grid geometry is shown in Table~\ref{lucietable}.  Dielectric constants are included as LXe 1.95, GXe 1.0, PTFE 2.1, UHMWPE 2.3. Applied grid voltages are assigned as relevant to WS2013 operations; voltages of the field-shaping (dodecagonal) rings between cathode and gate follow expectation given the resistor within the voltage dividing chain.  


\begin{table}[ht]
\setlength{\extrarowheight}{3pt}
\caption{Grid properties and voltages as relevant to the construction of the electric field model, including description of geometric simplifications.  \label{lucietable}}
\begin{center}
\begin{threeparttable}
\begin{tabularx}{\linewidth}{lSSSS[table-format = 4.1]lS}
\hline
\hline
Grid &   \multicolumn{1}{c}{$z$\tnote{\dag}}    &\multicolumn{1}{c}{Wire\diameter}  & \multicolumn{1}{c}{Pitch}   &\multicolumn{1}{c}{Angle}   &\multicolumn{1}{c}{Modeled}  & \multicolumn{1}{c}{HV}\\
     &  \multicolumn{1}{c}{[cm]}     &\multicolumn{1}{c}{[$\mu$m]} & \multicolumn{1}{c}{[mm]} & \multicolumn{1}{c}{[deg]} &  \multicolumn{1}{c}{as} & \multicolumn{1}{c}{[kV]}\\
\hline
Top shield    &58.6 & 50.8 & 5.00 & 135 & Absent & -1.0 \\
Anode         &54.9 & 28.4 & 0.25 & \multicolumn{1}{l}{{N/A}} &  Plane & 3.5 \\
Gate          &53.9 & 101.6 & 5.00 & 15 & \diameter0 wires & -1.5 \\
Cathode       &5.6  & 206.0 & 5.00 & 75 & \diameter0 wires & -10.0 \\
Bottom shield$\,\,$ &2.0  & 206.0 & 10.00 &15 & Plane & -2.0 \\
\hline
\hline
\end{tabularx}
\begin{tablenotes}
\item[\dag] $z$ is defined as vertical distance from the face of the bottom PMT array, accounting for thermal contraction as appropriate.
\end{tablenotes}
\end{threeparttable}
\end{center}
\end{table}

\begin{figure}[t!] \includegraphics[width=85mm]{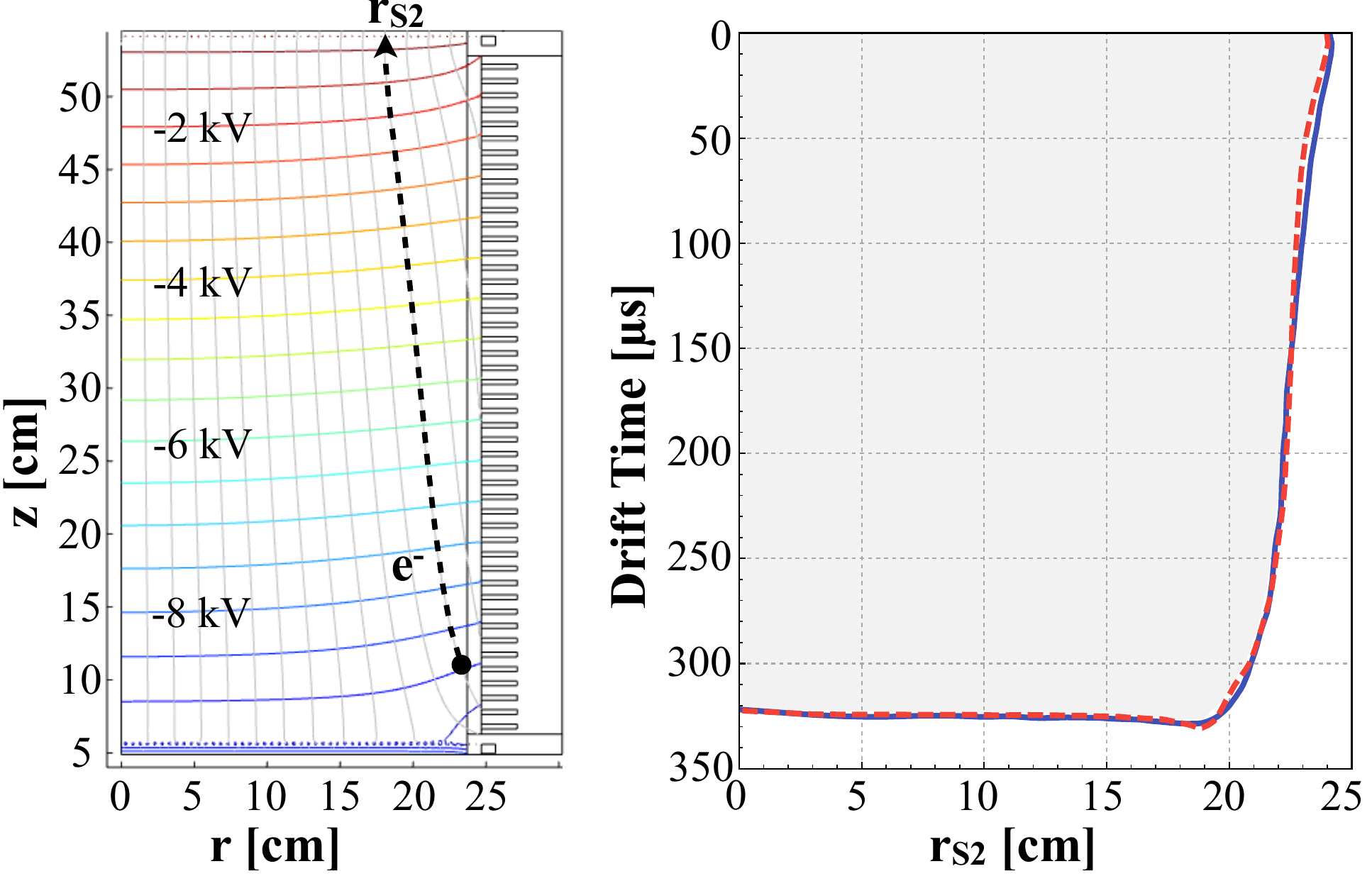} \caption{LEFT: A simplified 2D COMSOL Multiphysics model (for illustration only) shows electric field lines and equipotentials in the LUX detector under WS2013 conditions. A radially-inward component is seen, resulting from the geometry of the field cage and the grids.  RIGHT: a uniform distribution of electrons is drifted in the electric field model, and the edge of their resulting distribution in $t$\subStwo~and $r$\subStwo~is plotted here in solid blue.  A similarly defined edge can be drawn from the \krm data (dashed red), and the simulation and data can be seen to be consistent.  The edge is defined as event density contours, specifically as the contour at which the event density in \{$r$\subStwo$^2$, $t$\subStwo\} falls to 50\% of the average bulk value.} \label{fig:run3_field} \end{figure}

After solving for the electric field, field lines are used to simulate a uniform-activity dataset. Electron-like test particles follow the field lines to the liquid surface. The simulated electron drift velocity in LXe varies with electric field as in Ref.~\cite{driftvelocity}. The simulated drift time ($t$\subStwo) and $xy$ location of S2 light production ($x$\subStwo, $y$\subStwo) can be compared with real \krm data.  A simple 2D version of this 3D comparison space is illustrated in the right panel of Figure~\ref{fig:run3_field}.  We find excellent agreement between simulation and data.  It should be emphasized that no aspects of the field model are tuned to improve the level of agreement with data.

The slight curve seen in the reconstructed (S2) coordinates can be understood through inspection of the left panel of Figure~\ref{fig:run3_field}. This panel shows equipotentials and field lines from the simpler 2D (axially symmetric) model built for visualization purposes. When a \krm decay occurs at high radius just above the cathode plane, the liberated electrons follow the field lines shown and escape the liquid at a radius reduced by several cm compared to the interaction radius. The radial field component traces its origin primarily to the electrostatic transparency of the cathode and gate grids (both of 5.0~mm pitch).  This effect is strongest at high radii, producing a region of slightly reduced field above the cathode grid (created by upward leakage of the strong reverse-field region below the cathode) and a region of slightly enhanced field below the gate grid (created by downward leakage of the much higher above-gate field).

\section{Mapping S2 radius to vertex radius}\label{poscorr}

The field model could be employed as a mapping relating the observed S2 position and the true event vertex position, as \{$r$\subreal,~$\phi$\subreal,~$z$\subreal\}$~=~f($\{$r$\subStwo,~$\phi$\subStwo,~$t$\subStwo\}$)$.  However, we find that in the WS2013 field configuration, only the radial component of position required the construction of a detailed mapping function. The radial correction can be performed more precisely using the data alone, without relying on the accuracy of the field model and the electron drift simulation.  A data-driven method is possible and advantageous in WS2013 due to the small scale of the corrections, with the added benefit that it allows for the correction of all radial effects, including small-scale field inhomogeneities and systematic errors in the \{$x$\subStwo,~$y$\subStwo\} reconstruction algorithm.

\begin{figure}[t!]
\includegraphics[width=80mm]{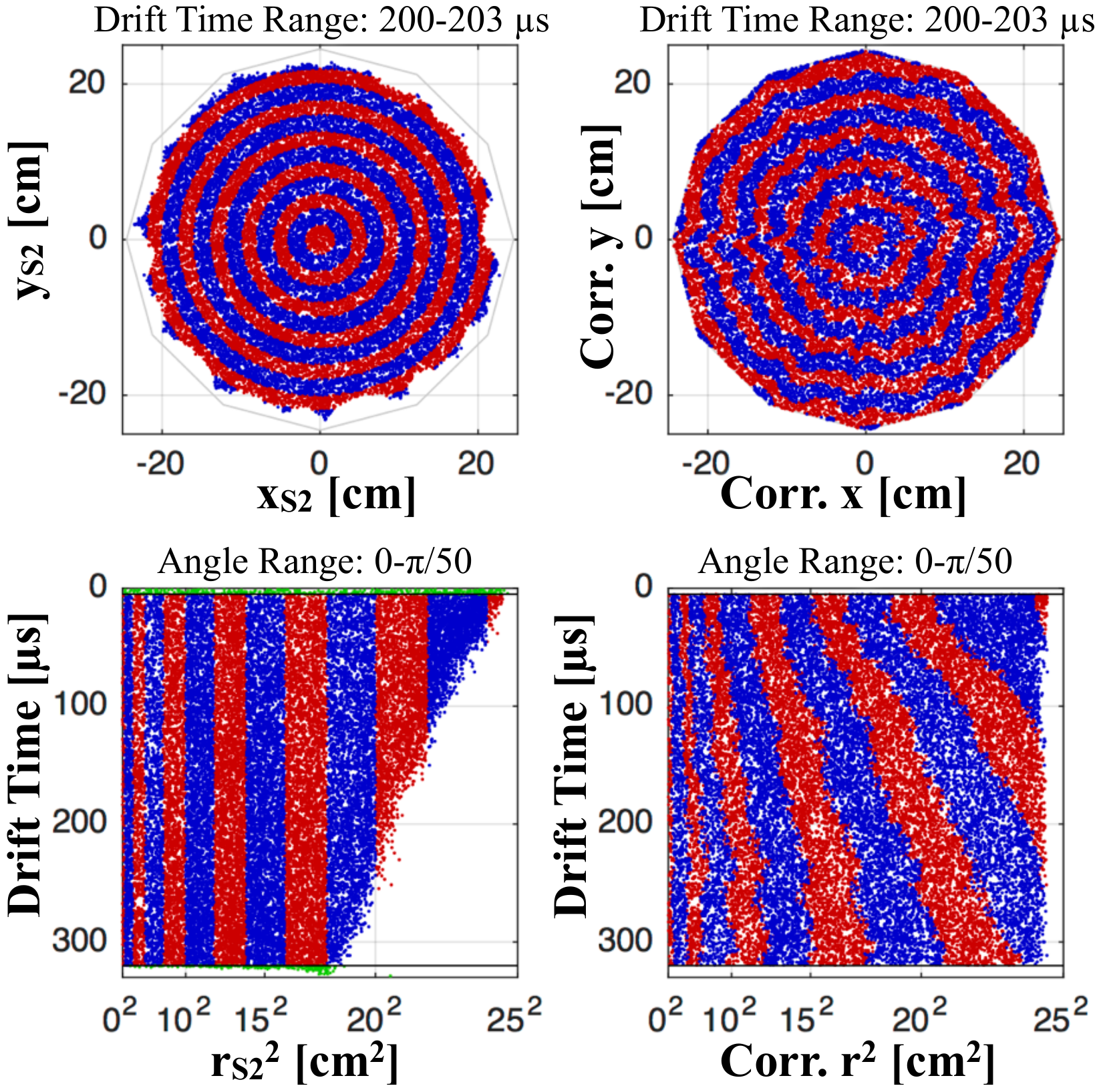}
\caption{An illustration of the effect of the radial position mapping between $r$\subStwo~ (left panels) and the resulting estimate of the true event radius (right panels).  The top panels show a thin slice of drift time, the bottom panels show a thin slice in angle.  In all panels, the same concentric selections in $r$\subStwo~ are highlighted (red and blue) to make the mapping visible.  Note that the mapping is only defined between 4 and 320~$\mu$s (events external to this range are green on the left, and not included on the right).  Note also that the use of squared radius in the lower panels exaggerates the scale of the effect.}
\label{fig:radialcorrection_method}
\end{figure}

\begin{figure}[t!]
\includegraphics[width=80mm]{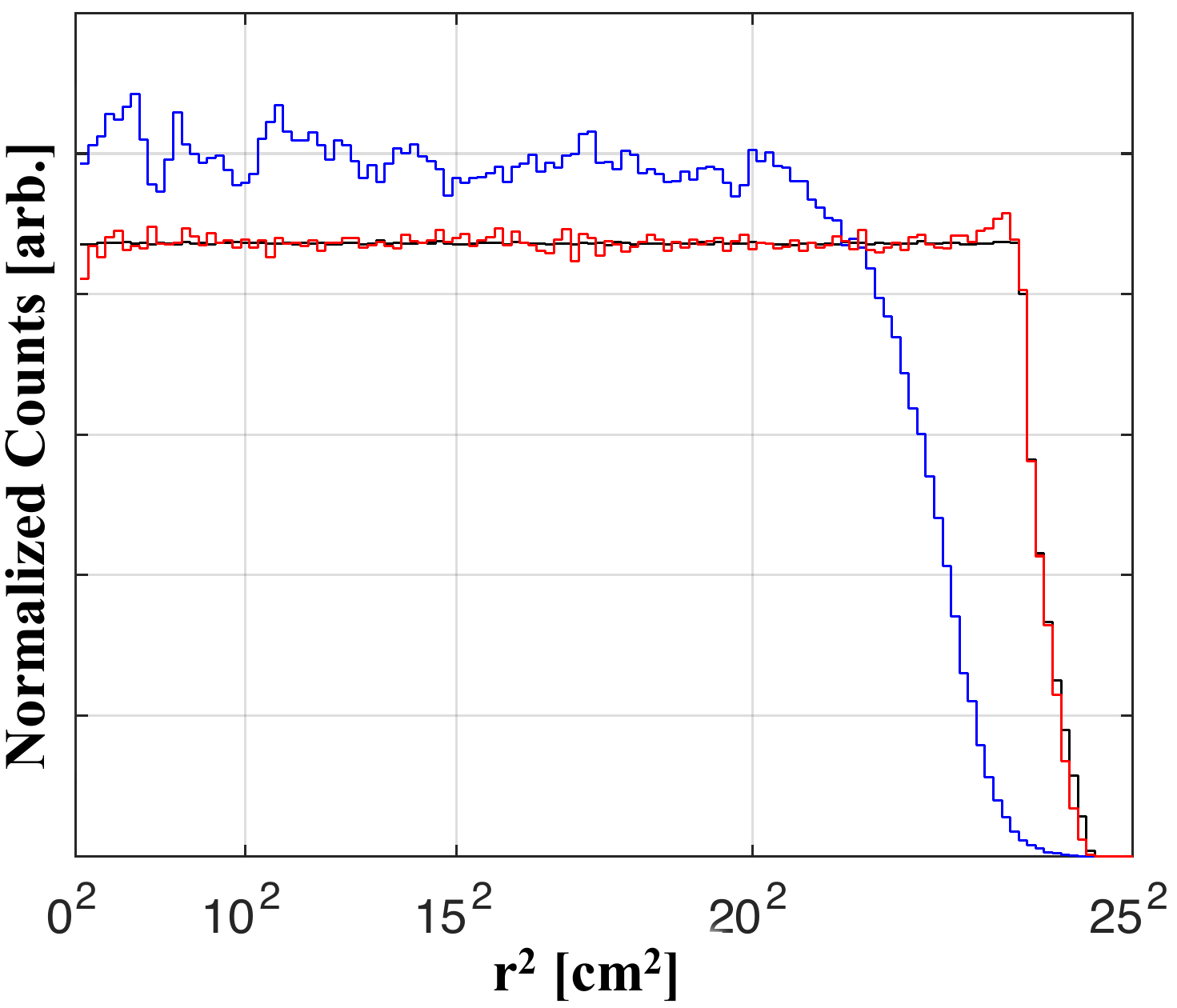}
\caption{The overall flattening effect of the radial position correction is illustrated.  A drift time selection is applied matching the WS2013 analysis. Three histograms in $r^2$ are shown: the $r$\subStwo$^2$ of a large sample of \krm events (blue), the $r^2$ distribution of those same \krm events after the radial correction procedure (red), and the distribution of events uniform in a dodecagonal prism in a toy Monte Carlo (black).  The \krm sample shown in the blue and red histograms of this plot combine a wide range of dates through WS2013 running.  The \{$x$\subStwo,~$y$\subStwo\} reconstruction algorithm used here differs slightly from that used in the published analyses.}
\label{fig:radialcorrection_result}
\end{figure}

The construction of a radius correction map relies on the uniform density of the \krm calibration events.  To ensure this uniform vertex density in real space, only data sufficiently long after activity injection ($\gtrsim$2~h) is employed in the map construction.  \krm events are grouped by vertex position into 11,520 wedge-shaped position selections:  A $t$\subStwo~range of 4 to 320~$\mu$s is divided into 32 $t$\subStwo~sections and 360 $\phi$\subStwo~sections.  These wedge selections of \krm events are not mutually exclusive, overlapping to the midpoint of neighboring selections in both $t$\subStwo~and $\phi$\subStwo.  Within a given wedge selection, the $r$\subStwo~distribution of \krm events is then `flattened' by shifting $r$\subStwo~values such that they are of equal spacing in $r^2$~(with maximum radius matching the appropriate dodecagonal radius at that $\phi$). Once each wedge selection region has received this treatment, the \krm event positions before and after the equal-spacing treatment are employed as a 3D linear interpolation mapping, as $r$\subreal$~=~f($\{$r$\subStwo,~$\phi$\subStwo,~$t$\subStwo\}$)$.  The application of this interpolative mapping function is illustrated in Figures~\ref{fig:radialcorrection_method} and \ref{fig:radialcorrection_result}.

Given that variation in the drift field over time will affect the position mapping function, temporal variation in the electric field is searched for using through Kolmogorov-Smirnov comparisons of \krm distributions on widely separated dates within the WS2013 and found to be consistent with no change. This allows the construction of a single WS2013 interpolative mapping function from a single large \krm injection from May 2013, supplying $1.5 \times 10^6$ selected \krm events.


\section{The position-dependent correction of S1 and S2 amplitudes}\label{areacorrections}

Detector efficiencies and gains may vary with position and time, requiring the construction of scintillation (S1) and ionization (S2) signal amplitude corrections. \krm events serve the role of `standard candles' to produce monoenergetic signals of uniform initial scintillation and ionization amplitudes, before efficiency and gain effects. The S1 and S2 cases receive somewhat distinct treatments, described below.

In the S1 case, detector efficiency variation is the result of a spatially-varying probability for a scintillation photon to strike a PMT window.  To map this efficiency, \krm data is binned in the 3D space of \{$x$\subStwo, $y$\subStwo, $t$\subStwo\}.  An average \krm S1 amplitude is found for each bin, and a 3D S1 correction map is constructed as the inverse of these \krm S1 amplitudes, normalized to the S1 amplitude at the detector center: \{0~cm, 0~cm, 159~$\mu$s\}.  The efficiency-correction map is then applied as a linear interpolation on the 3D grid.  Bin spacing of the \krm dataset was chosen such that each bin received $\sim$300 \krm events. It is observed that S1 correction maps vary negligibly with date, so a single large \krm injection provided the S1 correction map, subsequently applied to the full range of WS2013 data.

The S2 case is more complex.  A largely $z$-oriented efficiency variation dominates S2 variation, and results from electron capture on electronegative impurities during drift. During stable operation, the concentration of these impurities varies on a $\sim$week timescale.  An independent S2 amplitude variation, oriented purely in the $xy$ plane, results from three processes:  the efficiency of electron extraction across the liquid-gas boundary, and the efficiency of producing and then observing electroluminescence photons in the high-field gas region.  The extraction efficiency and electroluminescence yield can vary dependent on detector conditions such as pressure, liquid level (dependent on circulation flow rate), detector tilt, and electrostatic grid deflection.  

\begin{figure}[t!]
\includegraphics[width=80mm]{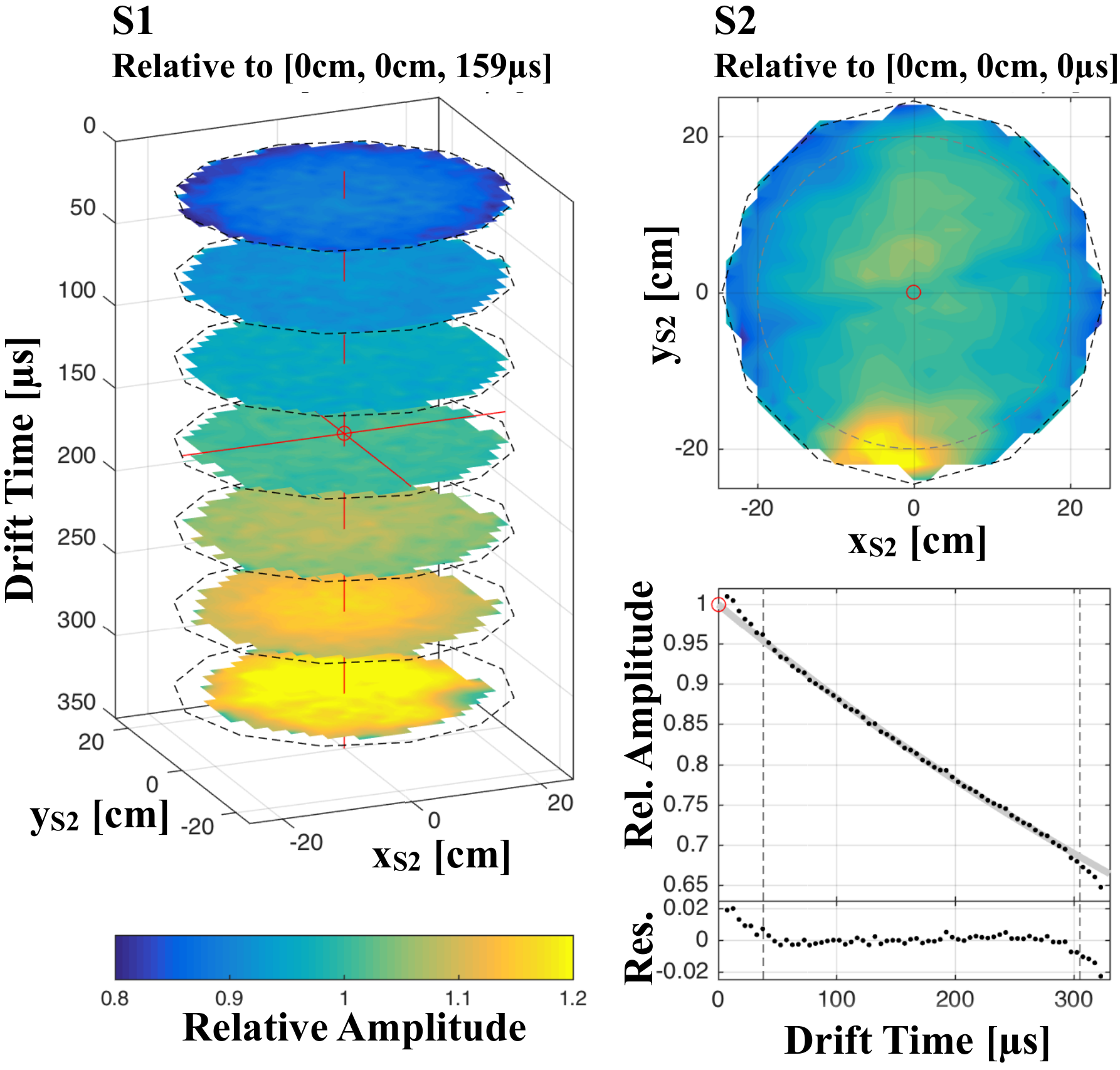}
\caption{Maps of relative \krm S1 and S2 amplitudes, as derived for an example date of May 10, 2013.  The 3D map of S1 amplitude is represented here as several slices in drift time.  To the right, the 2D $xy$ map of relative S2 amplitudes is shown (using the same colorscale), as is this date's 1D $z$ map, correcting for electron lifetime.  Note that the $z$-correction is applied using an exponential fit (here, $\tau_{e}$=805.2~$\mu$s, illustrated in gray). A residual for this fit is also shown.  Correction map normalization points are illustrated with red circles (the 3D center for S1, the top center for S2). The orientation of gate wires and gate region irregularities are visible in the S2 $xy$ correction map, and an inactive bottom-array PMT is apparent in the bottom of the S1 $xyz$ correction map.  Boundaries of the fiducial volume employed in \cite{lux2015} are indicated in the S2 plots by dashed gray lines.}
\label{fig:corrections}
\end{figure}

Two S2 correction maps are constructed, one for the $z$-dependent variation and one for the $xy$-dependent variation, and these maps are applied independently. The $z$-dependent S2 correction consists of a simple exponential function of $t$\subStwo, normalized to unity at the liquid surface (where electron lifetime has no effect on signal).  The single-valued $z$ correction is interpolated smoothly between measurements on \krm injection dates. It can be seen in Figure~\ref{fig:corrections} that while the exponential description of the $z$-dependent S2 correction describes the data well in the fiducial volume, it is an imperfect description at the extrema of the drift path, where the drift field deviates from its nearly constant bulk value.  This behavior at the extrema is consistent with impurities for which electron capture cross section decreases with field, a category including O$_2$~\cite{electronegative}.

The S2 correction $xy$ map is constructed by binning \krm data in an $xy$ grid and finding average S2 amplitudes for each 2D bin.  The $xy$ map is applied as a 2D linear interpolation of the inverse S2 amplitudes, normalized to $\{x$=0, $y$=0$\}$.  The $xy$ grid spacing was variable depending on \krm data sample size, binned such that each grid point represented $\sim$300 \krm events.  Temporal variation between consecutive $xy$ correction maps was much smaller than the $z$ correction variation; the $xy$ correction uses the nearest-in-time correction map.

\krm was injected weekly, a timescale set by variation in electron lifetime.  \isot{Rn}{222} decays (a constant, low-level background) supply independent verification of the electron lifetimes, and verification that the weekly \krm schedule was sufficiently finely-spaced.  Each \krm injection produces a typical sample size of $\sim10^5$ \krm decays. In the event of a sudden LXe purity change (such as a short circulation outage), data between the most recent \krm injection and the purity drop event are corrected assuming the last S2 correction map before purity change.  Data taken between a purity change and the first subsequent \krm are discarded. 

\begin{figure}[t!]
\includegraphics[width=80mm]{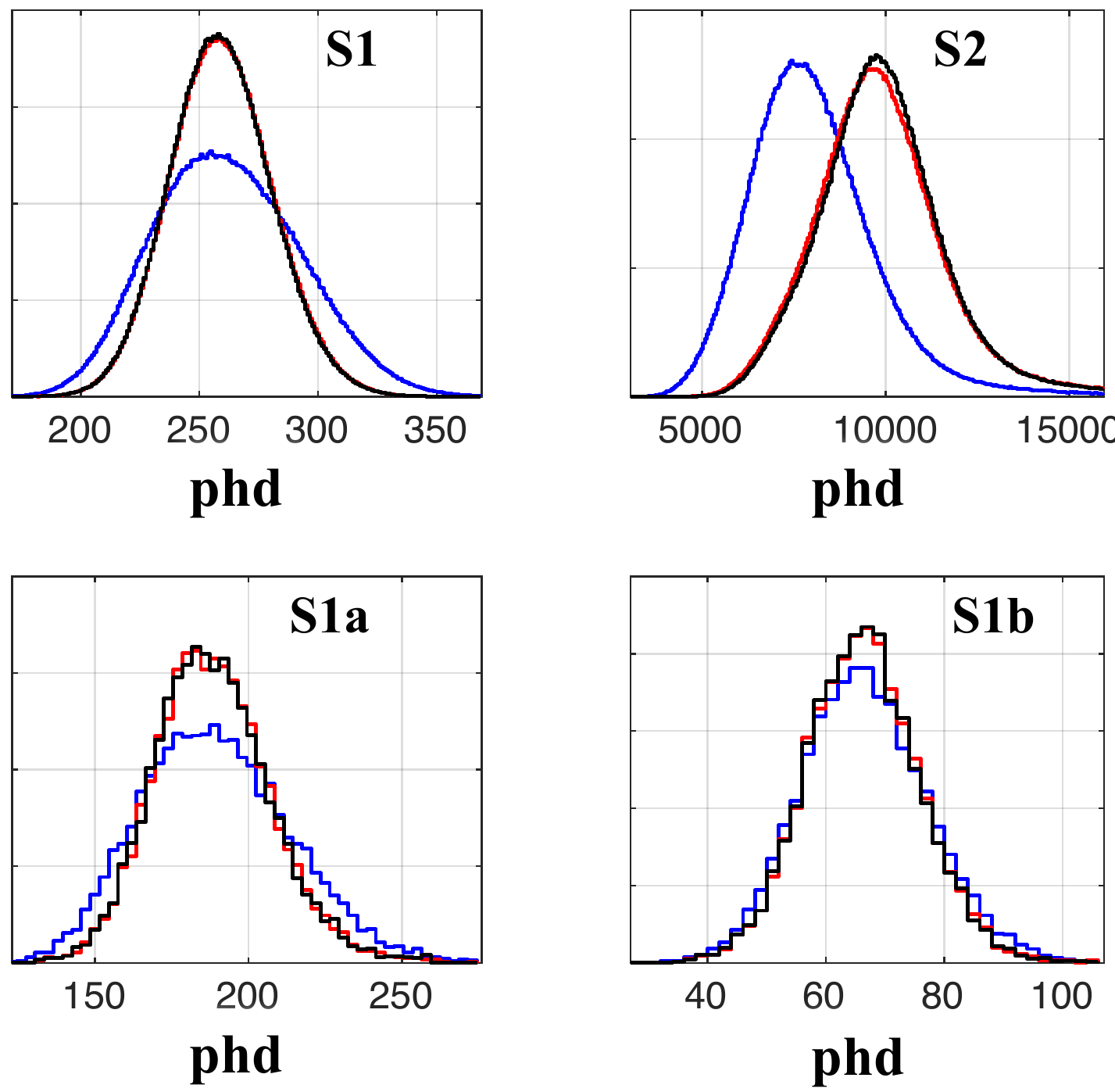}
\caption{The effect of applying the S1 and S2 amplitude correction maps is illustrated using the \krm S1 and S2 peaks themselves.  The starting distribution of uncorrected pulse area (measured in units of detected photons, phd) is shown in blue, a version corrected only in the $z$ direction is shown in red, and the final version corrected in all three spatial dimensions is shown in black.  The data here is a mixture of \krm data sets from a wide range of dates within WS2013 running, after applying the fiducial volume selection.  The top row shows the quantities used to create the corrections:  \krm decays for which the two S1 pulses are close enough together so as to be treated as a single pulse (t$_{\textrm{sep}}<$1.2~$\mu$s).  The lower row shows the individual S1 peaks when separation is achieved in the standard data treatment (t$_{\textrm{sep}}>$1.2~$\mu$s) and serve as a cross check of the S1 correction.}
\label{fig:corrected}
\end{figure}

As shown in Figure~\ref{fig:decay_diagram}, \krm decay proceeds through two transitions, separated by a 154~ns half-life.  Because the S2 signals are of 1.0--1.9~$\mu$s FWHM (depending on $z$ position), the two decay steps are merged in the S2 signal.  On the other hand, the S1 pulse width is short ($\sim$100~ns after filtering) such that a significant fraction of \krm decays exhibit separation into two S1 pulses, which we refer to by their ordering as S1a and S1b (32.1 and 9.4~keV, respectively).  It has been observed in \cite{baudis2013} that the S1b amplitude (and by implication the S1a+S1b summed amplitude) varies depending on the intervening time delay.  A short delay enhances electron-ion recombination in the second decay (S1b), increasing the resulting scintillation and thus boosting the S1b amplitude.  Because the S1a+S1b amplitude depends on the stochastic decay time between the two transitions, use of \krm S1 amplitude as a standard candle for calibration is only possible if one specifies and adheres to a consistent delay range at all positions.  Conversely, if a consistent delay range is used, the complexity of time delay amplitude variation can be ignored.  In the LUX WS2013 case, the summed S1 area is employed for S1 area corrections, and the transition time separation range is specified as 0 to 1200~ns.  These choices maximize useful calibration statistics.  S1 amplitudes in the separate S1a and S1b cases can be used as a cross check, as in the lower panels of Figure~\ref{fig:corrected}.

\begin{figure}[t!]
\includegraphics[width=80mm]{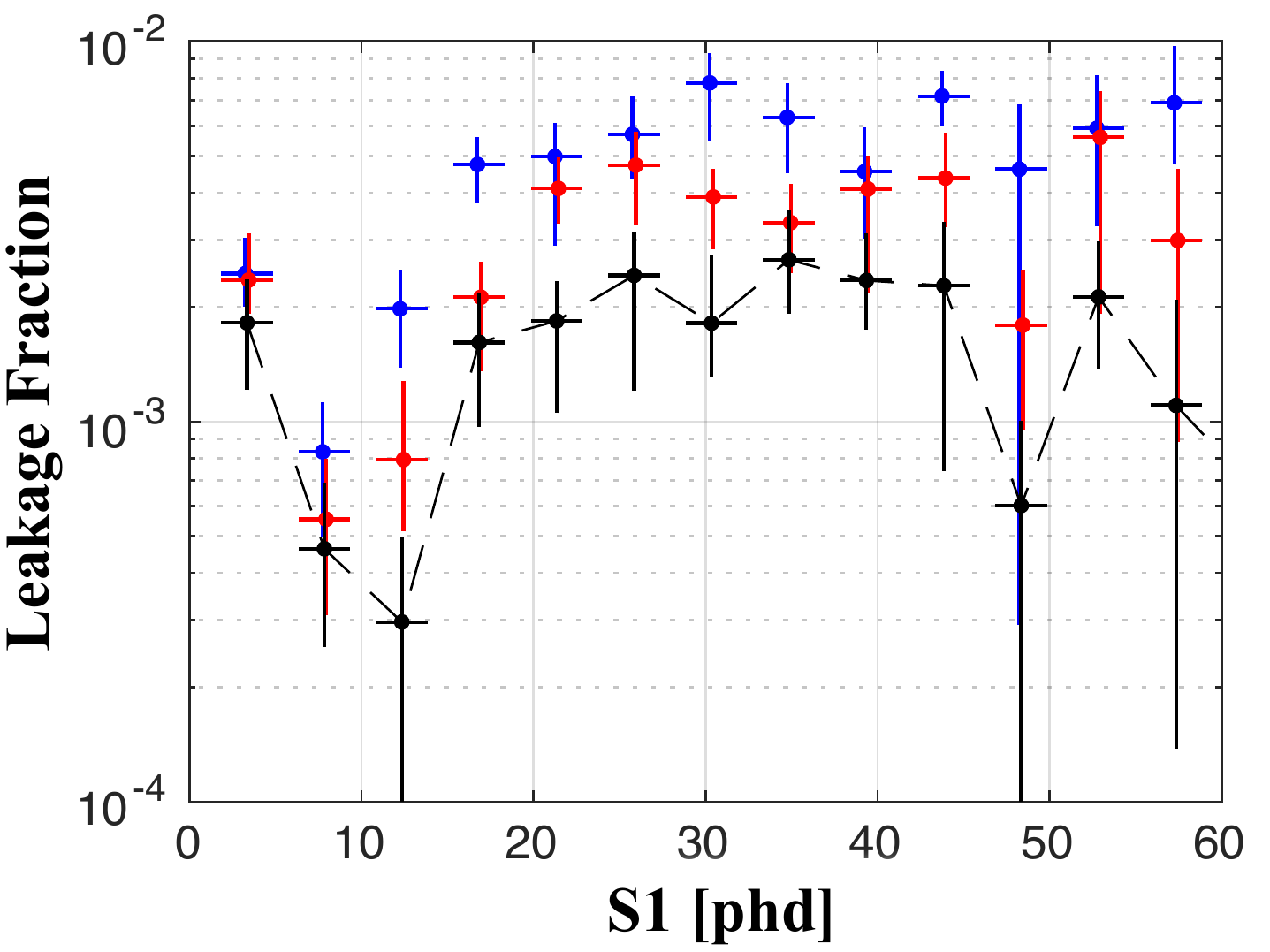}
\caption{A simple metric for background leakage fraction, the fraction of ER events falling below the NR S2/S1 mean, is shown binned in S1.  The ER sample used here is a $^3$H calibration, the NR sample is a calibration using a deuterium-deuterium (DD) neutron generator. Coloring matches Figure~\ref{fig:corrected}: blue denotes uncorrected areas, red denotes a correction only in drift time ($z$), and black denotes the full 3D correction. Uncertainties illustrated are statistical ($\sqrt[]{n}$) alone; S1 values are slightly offset to allow visibility.}
\label{fig:discrimination}
\end{figure}


The resolution and central value of the S1 and S2 peaks can be used to monitor the efficacy of amplitude correction maps. The resolution, $\sigma/\mu$, is calculated from Gaussian fits to the S1 and S2 amplitude distributions. For fiducial volume events in the largest \krm dataset (May 2013), the relative resolution of the combined 41.5~keV peak improves from 12.3\% to 8.1\% in S1 and from 19.3\% to 15.3\% in S2 after these corrections (see Figure~\ref{fig:corrected}). This S1 resolution improvement is typical of every data set, due to the stability of the position-dependent effects (e.g., photon mean free path, material surface reflectivities). On the other hand, the S2 improvement is highly dependent on the electron lifetime. For the largest \krm dataset, the lifetime is 750~$\mu$s, typical of WS2013 (which exhibits a range of lifetimes of 600--950~$\mu$s). We also look at the stability of \krm S1 and S2 central values over the course of the run after correction, and find that S1 varies by less than 0.6\%, and S2 varies by less than 2\%.  As expected, the S1 correction is of diminished importance for the resolution of small S1 amplitudes, where statistical fluctuations in photon number are of a similar or larger scale to the position-dependent variation. Indeed, the S1b (9.4~keV) peak resolution is the same with and without correction (15.4\% for events with S1a and S1b time separations of 1400 to 1600~ns).

The signal amplitude corrections enhance the electron recoil (ER) background rejection power of the S2/S1 discriminant quantity.  A simple metric of ER discrimination power is the fraction of ER events leaking past the nuclear recoil (NR) mean, using deuterium-deuterium neutron calibration data~\cite{dd} to define the NR mean and using tritium calibration data~\cite{tritium} to find the ER leakage.  This quantity is plotted as a function of S1 amplitude in Figure~\ref{fig:discrimination}, for varying levels of S1 and S2 amplitude correction of both ER and NR calibration datasets (no correction, $z$-only correction of both S1 and S2, and full 3D correction of both S1 and S2).  For S1 amplitudes of $>$10~phd, amplitude corrections are seen to enhance the discrimination power by a factor of $\sim$5.

Figure~\ref{fig:corrected} implies that a 3D correction represents only a marginal improvement over a $z$-only correction.  Important variations in $xy$ occur at high radius, outside the fiducial selection used in Figure~\ref{fig:corrected} or in the dark matter search analyses.  The improved discrimination when moving from $z$-only to full 3D correction parameters in Figure~\ref{fig:discrimination}, then, deserves some comment.  This improvement is partially a real change resulting from enhanced S1 and S2 resolution, but it is also partly an artifact of the NR calibration's specific and non-uniform position distribution (the NR calibration is performed using a narrow deuterium-deuterium neutron beam~\cite{dd}).  It so happens that the NR calibration distribution on the $xy$ plane is of very slightly enhanced S2 area, leading to an artificially-degraded discrimination measure before the $xy$ corrections are applied. 

The \krm calibrations of position and temporal variation lead directly to stronger ER discrimination and higher sensitivity to dark matter nuclear recoils, and were essential to the analyses published in \cite{lux2013}, \cite{lux2015}, and \cite{lux2017}.

\section{use of \krm amplitude ratios to map electric field amplitude }

The radial field component described in Section~\ref{fieldmodel} introduces a secondary effect: a drift field amplitude gradient in the $z$ direction.  Along the central axis, the field amplitude in WS2013 varies from $\sim$165~V/cm near the plane of the cathode grid to $\sim$205~V/cm near the plane of the gate grid.  A non-uniform electric field amplitude can produce a number of systematic effects, chief among them is a spatially-dependent fraction of electrons which recombine with ions.  A weaker field allows more recombination, enhancing the S1 signal and proportionally suppressing the S2 signal.  A stronger field has the inverse effect.  Field-dependence is minimal for low-energy electron recoils below 10~keV (where recombination is itself minimal) and increases above 10~keV~\cite{nest2011,tritium,luxyields}.  Electric field amplitude variations can also induce other systematic effects, including a spatially-dependent S1 pulse shape (through a varying recombination fraction as in the pulse amplitude case, see~\cite{LXePSD} and~\cite{luxPSD}) and a spatially-dependent electron lifetime (through field-dependent capture cross sections, as in ~\cite{electronegative}).  To the extent that these various systematics are important, a direct measure of local electric field amplitude in LXe is advantageous.


The S1 and S2 amplitude correction method described in Section~\ref{areacorrections} assumes \krm serves as a standard candle, and attributes all signal amplitude variation to detector efficiencies and gains.  The field dependence of initial photon and electron counts (before detector effects) relaxes the standard candle assumption, introducing a field-dependent variation that depends not only on event energy but on recoil type (ER or NR).  In the WS2013 science run described here, the scale of the field-dependence (at all energies and for both ER and NR) is estimated to be few-percent (following \cite{manalaysay2010, nest2011}), sub-dominant to other uncertainties, and is neglected.   

The small field dependence in \krm light and charge yields can be leveraged to construct a calibration quantity that varies with electric field amplitude alone.  When observably separated, the two S1 amplitudes of a \krm decay (at 32.1~keV and 9.4~keV) exhibit differing field dependence scales.  In fact, these two energies form a particularly convenient pair, in that 32.1~keV is well above the $\mathcal{O}(10~\textrm{keV})$ onset of significant field dependence, and 9.4~keV is just below.  The ratio of the two S1 amplitudes varies with field alone, since any S1 gain or efficiency effects affect both S1 amplitudes equally.

\begin{figure}[t!]
\includegraphics[width=80mm]{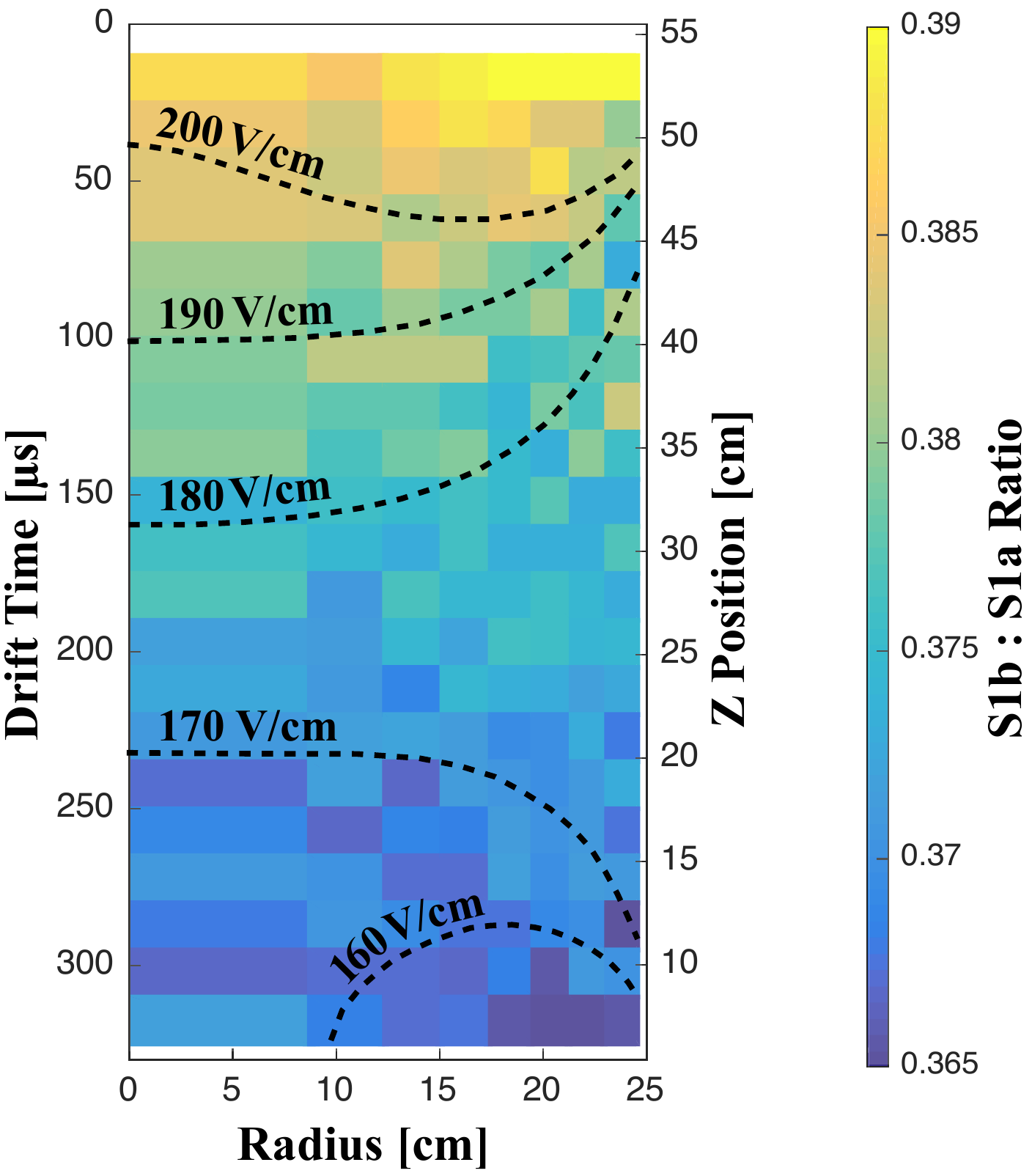}  
\caption{The \krm S1b:S1a ratio, plotted as average values within bins of drift time and (corrected) radius.  S1 amplitudes here are measured differently from other amplitudes in this work, by fitting an S1 pulse template.  A gradient in the S1b:S1a ratio is seen, matching the expected variation in field amplitude.  The high-field and low-field regions at high radius are also made visible.  Field amplitude contours are overlaid using a separate $z$ position axis in cm; correspondence to the drift time axis is accurate within a few percent.}
\label{fig:ratio_in_rz}
\end{figure}

The result is that the S1b:S1a ratio increases with the field.  Figure~\ref{fig:ratio_in_rz} shows a measurement of this ratio distribution in WS2013. Correspondence of this measured quantity with the field amplitude contours predicted by the field model of Section~\ref{fieldmodel} is clear.  The ratio measurement is statistically limited by the number of \krm decays for which the S1a and S1b pulses are measurably separated.  To maximize useful calibration statistics for this purpose, the two \krm S1 amplitudes are measured in a special data processing, employing a parameterized fit to estimate the individual amplitudes of slightly overlapping double-S1 traces.  This fit employs two instances of a single pulse template, fitting for four free parameters: two amplitudes and two pulse start times. This method allows S1a and S1b amplitude measurements down to a minimum separation time of 100~ns. This ratio technique for field amplitude measurement is of central importance in subsequent LUX analyses (as in \cite{lux2017}), for which the field variations in \krm recombination are larger and require careful treatment.

\section{Conclusions}

This first use of \krm in the calibration of a dark matter direct-detection experiment takes advantage of several key attributes of $^{83m}$Kr: a low-energy monoenergetic peak conveniently just above the energy region of interest, dispersible uniformly throughout the detector volume, with a convenient hours-scale decay time.  The monoenergetic signal enables a precise correction of S1 and S2 amplitudes for position-dependent efficiencies and gains, resulting in an enhanced ER background rejection ability.  The uniform spatial distribution enables a precise reconstruction of vertex position, enabling a well-defined fiducial volume selection.  This initial experience with \krm in a large-scale, operating dark matter experiment pointed the way towards unforeseen uses, such as the mapping of electric field amplitude variation within the TPC.  In subsequent LUX operations\cite{lux2017}, a buildup of electric charge on PTFE surfaces induced a time-varying and more significantly inhomogeneous electric field, requiring both a more complex procedure for the correction of positions and a more complex approach to S1 and S2 amplitude corrections.  These two $^{83\textrm{m}}$Kr-based calibration efforts, building on the LUX2013 experience described here, will be documented in two forthcoming papers~\cite{run4fieldpaper,run4krypcalpaper}.  Work is ongoing to make the most of \krm calibrations in current and future projects such as DarkSide~\cite{darkside2016}, XENON1T~\cite{XENON1T}, and LZ~\cite{LZ}.

\begin{acknowledgments}

This work was partially supported by the U.S. Department of Energy (DOE) under award numbers DE-AC02-05CH11231, DE-AC05-06OR23100, DE-AC52-07NA27344, DE-FG01-91ER40618, DE-FG02-08ER41549, DE-FG02-11ER41738, DE-FG02-91ER40674, DE-FG02-91ER40688, DE-FG02-95ER40917, DE-NA0000979, DE-SC0006605, DE-SC0010010, and DE-SC0015535; the U.S. National Science Foundation under award numbers PHY-0750671, PHY-0801536, PHY-1003660, PHY-1004661, PHY-1102470, PHY-1312561, PHY-1347449, PHY-1505868, and PHY-1636738; the Research Corporation grant RA0350; the Center for Ultra-low Background Experiments in the Dakotas (CUBED); and the South Dakota School of Mines and Technology (SDSMT). LIP-Coimbra acknowledges funding from Funda\c{c}\~{a}o para a Ci\^{e}ncia e a Tecnologia (FCT) through the project-grant PTDC/FIS-NUC/1525/2014. Imperial College and Brown University thank the UK Royal Society for travel funds under the International Exchange Scheme (IE120804). The UK groups acknowledge institutional support from Imperial College London, University College London and Edinburgh University, and from the Science \& Technology Facilities Council for PhD studentships ST/K502042/1 (AB), ST/K502406/1 (SS) and ST/M503538/1 (KY). The University of Edinburgh is a charitable body, registered in Scotland, with registration number SC005336.

This research was conducted using computational resources and services at the Center for Computation and Visualization, Brown University, and also the Yale Science Research Software Core. The \rb used in this research to produce \krm was supplied by the United States Department of Energy Office of Science by the Isotope Program in the Office of Nuclear Physics.

We thank Kevin Charbonneau at Yale University for his significant contributions to the dosing of the \krm generators used in this work.

We gratefully acknowledge the logistical and technical support and the access to laboratory infrastructure provided to us by SURF and its personnel at Lead, South Dakota. SURF was developed by the South Dakota Science and Technology Authority, with an important philanthropic donation from T. Denny Sanford, and is operated by Lawrence Berkeley National Laboratory for the Department of Energy, Office of High Energy Physics.

\end{acknowledgments}

\bibliography{Kr83m}

\end{document}